\documentclass[fontsize=11pt]{arxiv-article}
\usepackage{standalone}
\bibliography{References}

\usepackage{Definitions}
\RequirePackage{tikz}
\usetikzlibrary{calc}

\usepackage{pifont}
\usepackage{hyperref}
\usepackage{mathrsfs}

\allowdisplaybreaks

\newcommand{\OfficialTitle}{
  Convexity, large charge and the large-N phase diagram of the~$\varphi^4$~theory
}
\title{\setstretch{1.4}
  {\color{Thoughtless}\Huge\textbf{\dosserif\OfficialTitle}}
}

\hypersetup{pdfauthor={Rafael Moser, Domenico Orlando, Susanne Reffert},pdftitle={\OfficialTitle}}

\author{%
  \begin{minipage}{.97\linewidth}
    \vspace{1cm}
    \begin{center} \dosserif%
      {\small
      	\textbf{Rafael Moser}\textsuperscript{\ding{73}}, 
         \textbf{Domenico Orlando}\textsuperscript{\ding{72}\ding{73}} and
         \textbf{Susanne Reffert}\textsuperscript{\ding{73}} 
                  }
    \end{center}
    \vspace{1cm}
    \authorBlock{\ding{73}}{\dosserif{} Albert Einstein Center for Fundamental Physics\\
       Institute for Theoretical Physics, University of Bern,\\
       Sidlerstrasse 5, CH-3012 Bern, Switzerland}
      \authorBlock{\ding{72}}{\dosserif{} INFN sezione di Torino,\\
       via Pietro Giuria 1, 10125 Torino, Italy}
  \end{minipage}
}

\date{}

\begin{document}

\setstretch{1.2}

\numberwithin{equation}{section}

\begin{titlepage}

  \maketitle

  \thispagestyle{empty}

  \vfill\dosserif{}

  \abstract{\normalfont{}\noindent{}%
    In this note we discuss the phase space of the O(2N) vector model in the presence of a quadratic and a quartic interaction by writing the large-N effective potential using large charge methods in dimensions $2<D<4$ and $4<D<6$.
    Based on a simple discussion of the convexity properties of the grand potential, we find very different behavior in the two regimes: while in $2<D<4$, the theory is well-behaved, the model in $4<D<6$ leads to a complex CFT in the UV, consistently with earlier results. We also find a new metastable massive phase in the high-energy regime for the theory on the cylinder.
  }

\vfill

\end{titlepage}

\setstretch{1.2}

\tableofcontents

\section{Introduction}
\label{sec:intro}

In the study of strongly coupled systems it is often convenient to introduce control parameters with the goal of organizing perturbatively an analytic computation in an appropriate limit.
In the case of the \(O(2N)\) vector model, in the limit \(N \to \infty\) one obtains an expansion around a semiclassical description in terms of a collective field~\cite{stratonovich1957method,Hubbard:1959,Moshe:2003xn,Vasiliev:1981yc,Lang:1990ni,Petkou:1994ad,Petkou:1995vu}.
Remarkably, this semiclassical description captures the qualitative features of the quantum behavior of the system.
However, it is nonetheless useful to introduce yet another control parameter in order to extract physical information from this semiclassical description.
We choose to work in a sector of large global charge $Q$ here~\cite{Hellerman:2015nra}\footnote{For a review of the large charge expansion see~\cite{Gaume:2020bmp}.}, working in the double-scaling limit \(N \to \infty\), \(Q \to \infty\) with \(Q/N\) fixed~\cite{Alvarez-Gaume:2019biu,Orlando:2019hte,Giombi:2020enj,Dondi:2021buw}\footnote{Also other double-scaling limits, \emph{e.g.} involving the $\epsilon$ expansion have proved useful in the context of the large charge expansion, see~\cite{Watanabe:2019pdh,Badel:2019oxl,Badel:2019khk,Arias-Tamargo:2019xld,Antipin:2020abu,Antipin:2020rdw,Antipin:2021akb,Jack:2021lja,Jack:2021ypd,Jack:2021aui}.}.

In this work (and the companion paper~\cite{Orlando:2021usz}) we use this technology to derive the large-\(N\) effective potential without having to resort to Feynman-diagram techniques. The large-\(N\) effective potential of the \(O(2N)\) has been studied in a variety of works, starting from the seminal work of Coleman, Jackiw and Politzer~\cite{Coleman:1974jh}. 
Since we are interested in the consistency of the theory in various dimensions and want to calculate scaling dimensions via the state-operator correspondence, it is natural for us to study the effective potential in terms of the vector field, as done in~\cite{Appelquist:1981sf,Appelquist:1982vd}, instead of the Stratonovich collective field used in~\cite{Coleman:1974jh}.
Our observation is that these two effective potentials are related by a Legendre transform. In our fixed-charge analysis, the latter is identified with the grand potential.
In this way we can describe the phase diagram of the vector model in the presence of \(\phi^2\) and \(\phi^4\) operators in any dimension in a compact way, generalizing previous computations that typically concentrate on the behavior around the fixed points~\cite{Appelquist:1981sf,Appelquist:1982vd}. This goes far beyond previous applications of the large charge expansion, which has to date been mostly studied at the conformal point only, with few exceptions~\cite{Orlando:2019skh,Orlando:2020yii,Orlando:2020idm}.

Here we concentrate on the interval \(4 < D < 6\) in Euclidean space (see~\cite{Parisi:1975im,Fei:2014yja}).
Imposing the convexity of the effective potential, we find that the standard \(\phi^4\) model is \emph{not} \ac{uv} complete if we require unitarity,  consistently with recent observations~\cite{Giombi:2019upv,Giombi:2020enj,Antipin:2021jiw}.
We also present a possible completion in terms of a complexified effective potential, describing the flow to a non-unitary \ac{cft} for which we compute the (complex) dimensions of the lowest operators of given charge, finding complete agreement with the existing literature~\cite{Giombi:2020enj,Giombi:2019upv,Antipin:2021jiw,Arias-Tamargo:2020fow}.
This is reminiscent of complex \ac{cft}s obtained from the merging and annihilation of perturbative fixed points, corresponding to walking \ac{rg} flows and weakly first-order phase transitions~\cite{Sannino:2009za,Sannino:2012wy,Dymarsky:2005uh,Pomoni:2009joh,Kaplan:2009kr,Giombi:2015haa,Grabner:2017pgm,Gorbenko:2018ncu,Benini:2019dfy,Antipin:2020rdw}; however, in the case at hand, it seems that the mechanism of the generation of complex dimensions is different since it is related to instabilities of the theory.
Having access to the phase diagram, we also find a previously unknown metastable massive phase for the model on the cylinder \(\setR \times S^d\), much in the spirit of~\cite{Giombi:2019upv}, and we show that the \ac{cft} remains unstable also in the small-charge regime in which the conformal dimensions are real~\cite{Bekaert:2011cu,Bekaert:2012ux,Arias-Tamargo:2020fow,Giombi:2020enj,Antipin:2021jiw}.

More in general, having a non-perturbative description of the effective potential, we can relate the stability of the model to the convexity of its free energy and -- at criticality, via the state-operator correspondence -- to the convexity of the conformal dimension \(\Delta(Q)\) of the lowest operator of given charge.
This provides further evidence in favor of the recent proposal by Aharony and Palti related to the weak gravity conjecture~\cite{Aharony:2021mpc}, confirming the recent large-charge analysis in~\cite{Antipin:2021rsh}.

\bigskip

The outline of this paper is as follows.
In Section \refstring{sec:EFT-Legendre-convexity} we discuss classical arguments for the convexity of the effective potential. We then go on to apply this knowledge to the context of a theory at fixed charge $Q$. There we encounter two Legendre transforms: one that relates the effective potential to the grand potential and one that relates the grand potential to the free energy.
We discuss these relations qualitatively and discuss the issue of convexity.\\
In Section~\refstring{sec:phi4} we apply the acquired knowledge to the $\phi^4$-theory with a global $O(2N)$ symmetry in the large-$N$ limit. We discuss the theory and the effective potential for spacetime dimensions $2<D<4$ and $4<D<6$.
Using Legendre transforms we show that the behavior and the properties of the effective potential varies significantly between the two cases. We argue that, while the theory flows to a \ac{wf} fixed-point for $2<D<4$, in $4<D<6$ the $\phi^4$-theory is not unitary.
We discuss the case $D=5$ in detail and explicitly compute the effective potential as a Legendre transform of the known grand potential. We show that the effective potential exhibits a branch cut and becomes complex in the \ac{uv}. \\
In Section~\refstring{sec:critical} we investigate the strongly coupled fixed points. We use a perturbative definition of the Legendre transform to compute the scaling dimension in both the large $Q/N$ and the small $Q/N$ regime.
We first discuss $D=3$ and $D=5$ for both regimes and compare the resulting scaling dimensions.
We show that the results for $D=5$ are unsound: for large $Q/N$ the theory is non-unitary and for small $Q/N$ the free energy/scaling dimension appears to be non-convex in $Q$.
Meanwhile, for $D=3$, the scaling dimension is perfectly sound, convex and shows no violation of unitarity.
In addition, a resurgent analysis can be performed to interpolate between the small $Q/N$ and the large $Q/N$ regime~\cite{Dondi:2021buw}.
We then also discuss the spacetime dimensions $D= 4\pm\epsilon$ and $D= 6\pm\epsilon$ and compute the leading term of the scaling dimension.\\
In the Appendix we present the resurgent analysis for the theory in $D=5$ using the  techniques of~\cite{Dondi:2021buw}, extending from the two-sphere to higher dimensions.

\section{Effective actions, Legendre transforms and convexity}
\label{sec:EFT-Legendre-convexity}

\paragraph{Convexity of the effective potential.}

To start, let us review some classical arguments on the convexity of the effective potential~\cite{Iliopoulos:1974ur,Israel+2015,duncan2012conceptual}. Consider the theory of a complex scalar field \(\phi\), invariant under a \(U(1)\) symmetry that acts linearly, \(\phi \to e^{i \epsilon} \phi\).
We want to formulate the theory in terms of an effective potential.
Using the standard procedure we add a linear source term to the path integral and write the generating functional
\begin{equation}
  Z[J]   = \braket{0}_J = \frac{\int \DD{\phi} e^{-S[\phi] - \int J \phi \dd[D]{x}}}{\int \DD{\phi} e^{-S[\phi]}} .
\end{equation}
The logarithm of \(Z[J]\) is the connected generating functional
\begin{equation}
  W[J] = \log Z[J].
\end{equation}
It is convenient to consider the Legendre dual of \(W[J]\).
We define the classical field
\begin{equation}
  \phi_c = \fdv{W}{J}
\end{equation}
and the effective action
\begin{equation}
  \Gamma[\phi_c] = W^*[\phi_c] = \eval{J \phi_c - W[J]}_{J=J(\phi_c)} .
\end{equation}
Expanding \(\Gamma[\phi_c]\) around the value \(\phi_c = \text{const.}\) we find
\begin{equation}
  \Gamma[\phi_c]   = \int \dd[D]{x} \bqty{ V(\phi_c) +  Z(\phi_c) \del_\mu \phi_c^* \del_\mu \phi_c + \dots } ,
\end{equation}
where \(V\) is the effective potential.
The \(U(1)\) symmetry requires all the functions to depend only on the absolute value \(\abs{\phi_c}\), so that $V(\phi_c) = V(|\phi_c|)$.

As long the path integral has a positive definite measure, \(W[J]\) is a convex function.
This is in general true for the logarithm of the partition function as function of any parameter that enters linearly in the action.
The argument is based on Hölder's inequality: if \(\dd{\mu}\) is a positive measure, then for \(f\) and \(g\) positive functions one has
\begin{equation}
  \int \dd{\mu} f^\lambda g^{1-\lambda} \le \pqty{\int \dd{\mu} f}^\lambda \pqty{\int \dd{\mu} g}^{1 -\lambda} ,
\end{equation}
where $0\leq \lambda \leq 1$.
For a given field theory, we define the normalized measure
\begin{equation}
  \dd{\mu} = \frac{\DD{\phi} e^{-S[\Phi]}}{\int \DD{\phi} e^{-S[\phi]}} 
\end{equation}
and consider the expectation value
\begin{equation} 
\begin{aligned}
  Z[\alpha] &= \frac{1}{Z[0]} \ev{\exp[\int\dd[D]{x} \sum_i \alpha_i(x) f_i(\phi)]} 
  = \frac{1}{Z[0]}  \int \DD{\phi}  e^{-S[\Phi]} \exp[\int\dd[D]{x} \sum_i \alpha_i f_i(\phi)] \\
  &= \int \dd{\mu} \exp[\int\dd[D]{x} \sum_i \alpha_i f_i(\phi)] ,
\end{aligned}
\end{equation}
where the \(\alpha_i\) are functions of \(x\). For \(0< \lambda < 1\), take
\begin{equation}
  Z[\lambda \alpha + (1- \lambda) \beta] =  \int \dd{\mu} \exp[\int\dd[D]{x} \sum_i \pqty{\lambda \alpha_i + ( 1 - \lambda) \beta_i }f_i(\phi)].
\end{equation}
By the inequality,
\begin{equation} 
  Z[\lambda \alpha + (1- \lambda) \beta] \le \frac{ \bigg( \int \dd{\mu} \exp[\int\dd[D]{x} \sum_i \alpha_i f_i(\phi)] \bigg)^\lambda }{ \bigg( \int \dd{\mu} \exp[\int\dd[D]{x} \sum_i \beta_i f_i(\phi)]  \bigg)^{\lambda-1} }
  = Z[\alpha]^\lambda Z[\beta]^{1-\lambda} ,
\end{equation}
and taking the logarithm,
\begin{equation}
  \log(Z[\lambda \alpha + (1- \lambda) \beta] )  \le \lambda \log(Z[\alpha]) + ( 1 - \lambda) \log(Z[\beta]) .
\end{equation}

The Legendre transform of a convex function is convex.
It follows that \(\Gamma[\phi_c]\) is convex and since this property has to remain true for constant \(\phi_c\), also the effective potential \(V(\phi_c)\) must be a convex function.\footnote{This is true also in finite volume. For strongly-coupled systems that are studied numerically, it is convenient to introduce another quantity, the constraint effective potential $U$:
\begin{equation}
	e^{-U(\phi)} = \int \DD{\hat\phi} \delta \left(\frac{1}{V}\int \dd[D]{x} \pqty{\phi - \hat \phi} \right) e^{-S[\hat \phi]},
\end{equation}
which is related to the effective potential by a Legendre transform.
$U(\phi)$ is in general not convex but coincides with $V(\phi_c)$ in the decompactification limit~\cite{ORaifeartaigh:1986axd}.}

\paragraph{Fixed charge and Legendre transforms.}

We want to study our system in a sector of fixed charge.
We start with an effective action with a canonical kinetic term%
\begin{equation}
  \Gamma[\phi_c] = \int \dd[D]{x} \bqty{ \partial_\mu \phi_c^* \partial_\mu \phi_c - V(\abs{\phi_c})}.
\end{equation}
Then the conserved \(U(1)\) charge has the form
\begin{equation}
  Q = i \int \dd[D-1]{x} \pqty{ \dot \phi_c^*\phi_c - \phi_c^* \dot \phi_c} .
\end{equation}
To study the system at fixed charge, we make the ansatz\footnote{The imaginary unit in the argument is  due to the Wick rotation.}
\begin{align}
	\abs{\phi_c} &= \Phi, & \arg(\phi_c) &= -im\tau  ,
\end{align}
which corresponds to a ground state with fixed chemical potential.\footnote{This ansatz implies spontaneous symmetry breaking that can also happen in finite volume and finite $N$ in the limit of large charge (see the discussion in Appendix~B in~\cite{Gaume:2020bmp}).}
To compute the energy of the ground state as a function of \(Q\), we can solve the problem in two steps.
First we eliminate \(\Phi\) using the \ac{eom} for the radial mode,
\begin{equation}
  \dv{(\Phi^2)} \bqty{ m^2 \Phi^2 - V(\Phi)} = m^2 - \dv{V}{(\Phi^2)} = 0 ,
\end{equation}
to write the \ac{vev} of the Lagrangian as function of \(m^2\) alone,
\begin{equation}
  \label{eq:grand-potential-from-V}
  \omega(m) = \eval{ \Phi^2 m^2 - V(\Phi)}_{\Phi = \Phi(m)} .
\end{equation}
Then we write the corresponding energy density \(f\) using the momentum associated to \(m\) (\emph{i.e.} the charge density \(\rho\)):
\begin{align}
  \rho &= \fdv{\omega}{m} = 2 m \Phi^2(m) ,\\ \label{eq:EnergyDensityAsLegendreTransform}
  f(\rho) &=  \bqty{ \rho m - \omega(m)}_{m = m(\rho)}.
\end{align}
An effective action with canonical kinetic term clearly does not represent the general case. 
In the  $O(2N)$ model at leading order in \(N\) the effective action in terms of the fundamental field $\phi$ is non-local (see~\cite{Appelquist:1982vd,Townsend:1975kh,Townsend:1976sy}), but when evaluated on our ansatz it still takes the form in Eq.~\eqref{eq:grand-potential-from-V}, which is what we base our analysis on.

It is easy to recognize two Legendre transforms.
If we introduce the notation
\begin{align}
  x&= \abs{\phi_c}^2, & \Upsilon(x) &= V(\sqrt{x}), & y&= m^2 , & \varpi (y) &= \omega(\sqrt{y}) ,
\end{align}
the chain of transformations becomes
\begin{equation}%
\label{eq:Legendre-chain}
  V(\phi_c)= \Upsilon(\abs{\phi_c}^2) \to \Upsilon^*(m^2)= \varpi (m^2)= \omega(m) \to \omega^*(\rho)= f(\rho) .
\end{equation}
For convex functions, the direction of the arrows can be interchanged, since the Legendre transform is an involution, and we can, for example, compute the effective potential from the knowledge of \(\omega(m)\), as we will do in the following.

The minimization condition derived above admits in general complex solutions.
There are two possible stances we can take:
\begin{itemize}
	\item We define the Legendre transform as 
	\begin{equation}
	\label{eq:sup-def-Leg}
	  f^*(y) = \sup_x( x y - f(x)) .
	\end{equation}
	In this case, $f^*(y)$ takes values on the extended real line $\setR \cup \set{\pm\infty}$ and the result is always real.
	\item We use the naive definition as seen above which leads in general to multivalued complex functions.
\end{itemize}
The two definitions coincide as long as $f(x)$ is smooth and convex.
In this case, the Legendre transform is an involution and preserves convexity.
In general, the definition with the supremum always leads to a convex function.
While in the context of classical thermodynamics it is clear that the supremum definition is the physical one~\cite{Israel+2015}, the situation is less clear-cut for quantum systems, since also complex saddles of the path integral have a meaning, as exemplified by resurgence theory~\cite{Cherman:2014ofa}. 

As discussed above, of all the functions that we have defined only \(V(\phi_c)\) is always convex for a unitary theory.
In the following we will use this property as a \emph{necessary condition} for the unitarity of a given theory.
Note however that the convexity of \(V\) and the convexity of \(\Upsilon\) are related in a non-trivial way:
\begin{equation}
	\dv[2]{\Upsilon}{y} = \frac{1}{4\phi_c^2}\pqty{\dv[2]{V}{\phi_c}-\frac{1}{\phi_c}\dv{V}{\phi_c}}.
\end{equation} 
For large values of $\phi_c$, the effective potential has to be an increasing function, therefore $\Upsilon$ does not have to be convex in general.

\section{$\phi^4$ theory in $D<6$ dimensions}
\label{sec:phi4}

In this section we apply our general considerations to the case of the \(O(2N)\)-symmetric \(\phi^4\) model in \(D\) spacetime dimensions in the limit of large \(N\).
In this case it is possible to compute directly the grand potential \(\omega(m)\) that appeared in Eq.~\eqref{eq:grand-potential-from-V} and use it to derive the effective potential.

We start with the action for $N$ complex scalar fields $\phi_i$ on $\setR_t \times\mathcal{M}$,
\begin{align}\label{eq:SThetaphi4}
    S[\phi_i] = - \int \dd{\tau}\dd{\mathcal{M}} \left[g^{\mu \nu}(\del_\mu \phi_i)^\dag (\del_\nu \phi_i) - r(\phi_i^\dag\phi_i) - \frac{u}{2N}(\phi_i^\dag\phi_i)^2\right] .
\end{align}
Depending on \(D\), the quartic term is either relevant or irrelevant (from the \emph{p.o.v.} of the free theory):
\begin{itemize}
\item For \(2 < D < 4\), the operator is relevant. If we fine-tune \(r = R(D-2)/(4(D-1))\) (the conformal coupling), the theory flows from a free theory in the \ac{uv} for \(u = 0\) to a strongly-coupled \ac{cft} for \(u \to \infty\) (the \ac{wf} fixed point).
\item For \(4 < D < 6\), the operator is irrelevant. If we fine-tune \(r\), the expectation is that the flow connects an \ac{ir} free fixed point at \(u =0 \) to a strongly-coupled \ac{cft} in the \ac{ir} for \(u \to \infty\).
  In the following we will show that this latter theory is not unitary.
\end{itemize}

Following~\cite{Alvarez-Gaume:2019biu}, we want to compute the free energy of the model restricted to the completely symmetric representation, which corresponds to the energy of the homogeneous ground state in the sector of total fixed charge \(Q\)~\cite{Alvarez-Gaume:2016vff,Antipin:2020abu}.
In the double-scaling limit of large \(N\) and large $Q$ with $Q/N$ fixed (which can be large or small), this free energy is the Legendre transform of the grand potential $\Omega(m)= V_{\mathcal{M}} \omega(m)$, whose leading-order behavior can be computed formally generalizing Stratonovich's construction. As shown in~\cite{Alvarez-Gaume:2019biu,Orlando:2021usz},
\begin{equation}\label{eq:omega-general}
  \omega(m) = (2N) \bqty{ -\frac{1}{2V_{\mathcal{M}}} \zeta(-\sfrac{1}{2} \mid \mathcal{M}, m) + \frac{\pqty{m^2 - r}^2}{4u} } ,
\end{equation}
where \(\zeta(s \mid \mathcal{M}, m) \) is the zeta function for the operator \((\Laplacian_\mathcal{M}{} - m^2)\).

Since we want to write the effective action in flat space, we can use the zeta function on a torus of side \(L\).
It is convenient to write \(\zeta(s \mid T^{D-1}, m)\) as a Mellin integral,
\begin{equation}
  \zeta(s \mid T^{D-1}, m) = \frac{1}{\Gamma(s)} \int_0^\infty \frac{\dd{t}}{t} t^s e^{-m^2 t} \Tr[e^{t \Laplacian{} }],
\end{equation}
and the heat kernel in terms of Weyl's asymptotic expansion,
\begin{equation}
  \Tr[e^{t \Laplacian}] = \frac{V_{T^{D-1}}}{(4\pi t)^{(D-1)/2}} + \order{e^{-L^2/(4t)}}
\end{equation}
so that
\begin{equation}
  \zeta(s \mid T^{D-1}, m) = \frac{V_{T^{D-1}} \Gamma(s - \frac{D-1}{2} )}{(4\pi)^{(D-1)/2} \Gamma(s)} m^{D-1-2s}.
\end{equation}
The grand potential \(\omega\) is then given by
\begin{equation}
  \omega(m) = (2N) \bqty{ \frac{  \Gamma(-\frac{D}{2} )}{2 (4 \pi)^{D/2}} m^D + \frac{(m^2-r)^2}{4u}} .
\end{equation}
This function contains all the information about the leading-\(N\) behavior of the phase diagram of the vector model with quadratic and quartic operators. If we fine-tune $r$ to the conformal coupling, it describes the flow that joins the Gaussian (\(u \to 0\)) and the strongly coupled (\(u \to \infty\)) fixed points.

\begin{figure}
  \centering
  \includestandalone[mode=image]{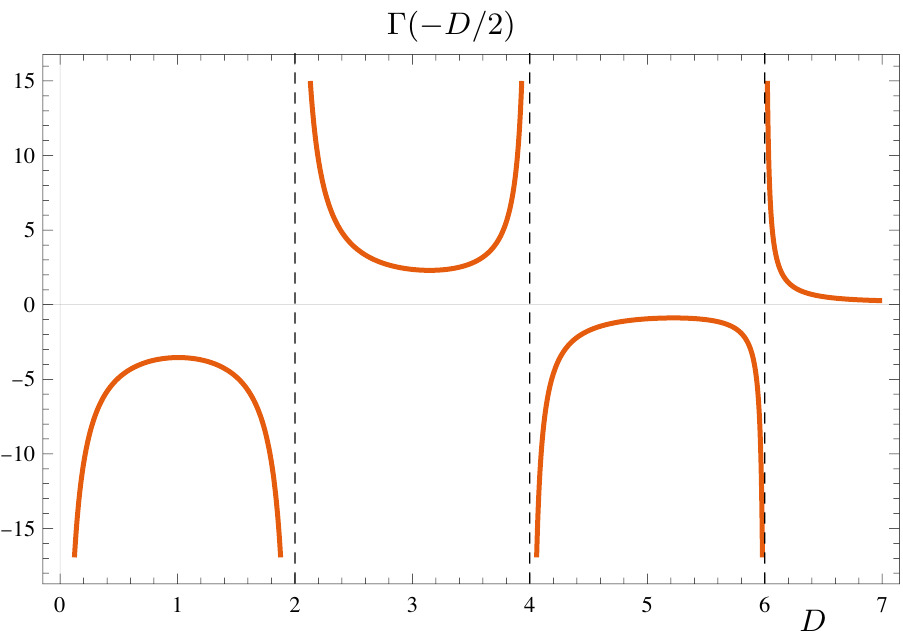}
  \caption{The leading term in the grand potential on the torus \(\propto \Gamma(-D/2)\) for \(0<D<7\).
  The function is positive for \(2 < D <4\), \(6 < D < 8\), etc.}
  \label{fig:Gamma-D}
\end{figure}

The only non-trivial function is the coefficient \(\Gamma(-D/2)\) of the first term.
It is positive if \(4n-2< D < 4n \), negative if \(4n < D <4n+2\) and diverges for \(D\) even (see Figure~\ref{fig:Gamma-D}).
This is consistent with the fact that there is no 
\ac{wf} point for even dimensions.

Starting from the grand potential we can use the technology developed in the previous section to write an effective potential \(V(\phi_c)\) for the vector model in terms of \(2N\) classical fields \(\phi_c^a\).
\(V(\phi_c)\) is determined by the condition that at fixed charge it must reproduce the physics in the double-scaling limit described by $\omega$ in Eq.~\eqref{eq:omega-general}~\cite{Orlando:2021usz}.
While $\omega$ was obtained from a fixed-charge computation, \(V(\phi_c)\) is valid for any value of the charge.
The \(O(2N)\) symmetry requires that the potential depends only on the invariant combination \(\phi_c^a \phi_c^a\), that by a slight abuse of notation we will indicate in the following as \(\phi_c^2\):
\begin{equation}
  \phi_c^2 = \sum_{i =1}^{2N} \phi_c^a \phi_c^a .
\end{equation}

As we have seen in Section~\ref{sec:EFT-Legendre-convexity}, the effective potential has to be a convex function.
Following the chain of transformations in Eq.~\eqref{eq:Legendre-chain} we can turn this into a consistency condition on the convexity of \(\varpi\).
To identify a possible flex in $\varpi$, we compute the second derivative and find:
\begin{equation}
  \varpi ''(y) = \frac{D (D-2) N}{4 (4 \pi)^{D/2}} \Gamma(-D/2) y^{(D -4)/2} + \frac{N}{u},
\end{equation}
which has a zero for positive \(y\) only when \(\Gamma(-D/2) < 0\). In Figure~\ref{fig:Gamma-D}, $\Gamma(-D/2)$ is plotted and we see that it is positive in certain intervals of the dimension and negative in others. We focus here on the intervals $2<D<4$ and $4<D<6$.

\subsection{$2<D<4$}
\label{sec:2to4}

For \(2<D<4\) the gamma function is positive, \(\Gamma(-D/2)>0\). It follows that \(\varpi\) is convex for all values of \(y\).
All the Legendre transforms are well-defined, and the theory fulfills our necessary condition for unitarity.
As shown in~\cite{Orlando:2021usz}, in \(D = 3\) we can compute the effective potential explicitly,
\begin{equation}
\begin{multlined}
  V(\phi_c) = \frac{N u^3}{3 \times 2^{10} \pi^4} \bigg( 1 + 96 \pi^2 \pqty{ \frac{\phi_c^2}{N u} + \frac{r}{u^2}  } + 1536 \pi^4 \pqty{ \frac{\phi_c^2}{N u} + \frac{r}{u^2}}^2 \\
   - \pqty{1 + 64 \pi^2 \pqty{ \frac{\phi_c^2}{N u} + \frac{r}{u^2}  }}^{3/2} \bigg) ,
  \end{multlined}
\end{equation}
In the critical \(r = 0 \) case this result had originally been found in~\cite{Appelquist:1982vd} from a resummation of infinitely many Feynman diagrams. 

This expression becomes more transparent if we look at the limits of small \(u\) and large \(u\), for example at \(r = 0\).
\begin{itemize}
\item For small values of \(u\), the effective potential reproduces the standard large-N loop expansion around the Gaussian fixed point~\cite{Appelquist:1982vd},
  \begin{equation}
    V(\phi_c) = \frac{u}{2N} \phi_c^4 \bqty{ 1 - \frac{1}{3\pi} \frac{\sqrt{N u}}{\phi_c} + \order{u}  } .
  \end{equation}
\item For large values of \(u\) we have a perturbative expansion in \(1/u\) around the \ac{wf} point,
  \begin{equation}
    V(\phi_c) = \frac{16 \pi^2}{3 N^2} \phi_c^6 \bqty{ 1 - 24 \pi^2 \frac{\phi_c^2}{N u} + \order{u^{-2}} } .
  \end{equation}
\end{itemize}

In $2<D<4$, the condition of convexity is always fulfilled. This is however only a necessary condition for unitarity.
In fact, it appears that for \(D = 4 - \epsilon\) the theory is not unitary~\cite{Hogervorst:2015akt} even though the effective potential is convex.

\subsection{$4<D<6$}
\label{sec:4to6}

For \(4 < D < 6\), the gamma function is negative, \(\Gamma(-D/2) < 0\), so  \(\varpi \) and has a flex for positive values of \(y\):
\begin{align}
  \varpi ''(\bar y) &= 0 & \text{for }&& \bar y^{(D-4)/2} &= \frac{4 (4\pi)^{D/2}}{u D (D-2) \abs{\Gamma(-D/2)}} .
\end{align}

\begin{figure}
  \centering
  \includestandalone[mode=image]{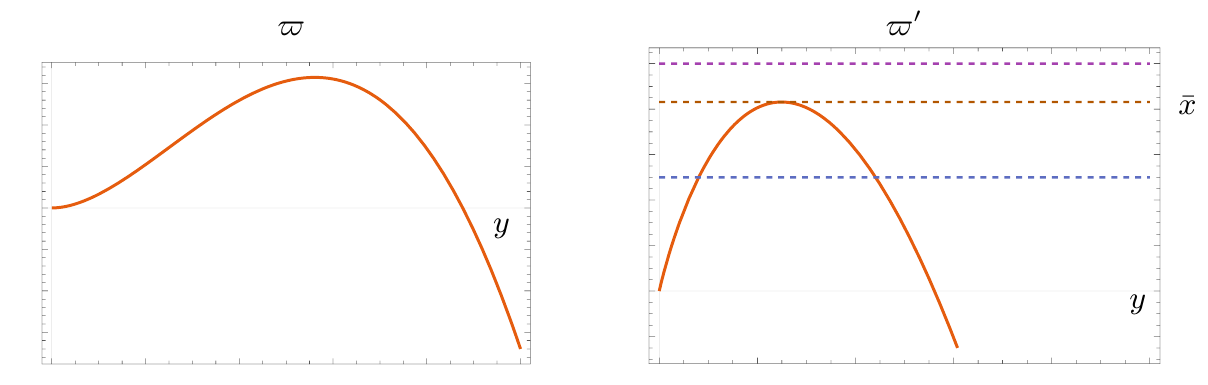}
  \caption{The function \(\varpi (y)\) (left) and its first derivative \(\varpi '(y)\) (right) on flat space (the torus).
    The function \(\varpi (y)\) has a flex at \(y = \bar y\) that separates a convex region for \(y < \bar y\) from a concave one.  The maximization condition \(x = \varpi '(y)\) admits two solutions for \(x < \bar x = \varpi '(\bar y)\) and no real solutions for \(x > \bar x\); the physical branch corresponds to the first intersection (right).}
  \label{fig:Omega-5d}
\end{figure}

The flex separates a convex region for small values of \(u\) around the free \ac{ir} fixed point from a concave region for large \(u\), around the conjectural strongly coupled \ac{uv} fixed point (Figure~\ref{fig:Omega-5d})).
Using  \(x = \varpi '(y)\) we can express the position of the flex in terms of \(\phi_{c}\):
\begin{equation}
  \bar x = \bar \phi_{c}^2 = \frac{\pqty{D - 4}(4\pi)^{D/(D-4) } N}{\pqty{\frac{D}{4} \abs{\Gamma(-\frac{D}{2} )}}^{2/(D-4)}  \pqty{D - 2}^{(D-2)/(D-4)}}  u^{-(D-2)/(D-4)} - \frac{N r}{u} .
\end{equation}
In this situation, we need to use the supremum definition of Legendre transform Eq.~\eqref{eq:sup-def-Leg}:
\begin{equation}
  \Upsilon(x) = \sup_{y > 0} (x y - \varpi (y)) .
\end{equation}
For \(x < \bar x\), the supremum is obtained by differentiating the argument at fixed \(x\) and expressing \(y\) as function of \(x\):
\begin{equation}
  x = \varpi '(y) .
\end{equation}
For \(x > \bar x\), on the other hand, there is no value of \(y\) such that \(x = \varpi '(y)\) (see Figure~\ref{fig:Omega-5d}).
The supremum is then obtained by minimizing \(\varpi \), but this function is not bounded below, so the supremum is \(+ \infty\).%
\footnote{Technically we are using the notion of convex conjugate, defined on the extended real line \(\setR \cup \set{\pm \infty}\)).}

\begin{figure}
  \centering
  \includestandalone[mode=image]{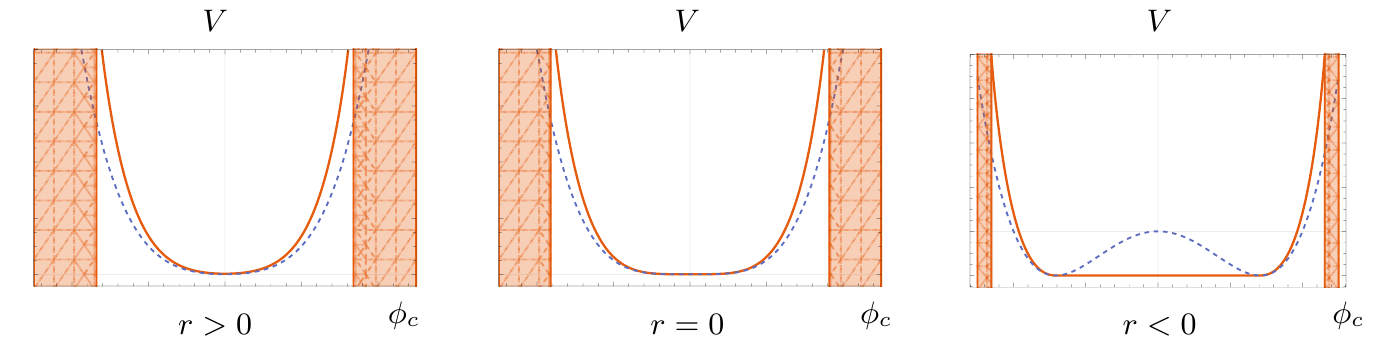}
  \caption{Effective potential in five dimensions: unbroken (\(r>0\)), critical (\(r=0\)) and broken phase (\(r<0\)).
    The dotted lines represent the tree-level potential, the red lines represent the effective potential that is always convex.
    In the shaded region the effective potential is infinite, signaling the breakdown of the \ac{eft} which requires a \ac{uv} completion.
   }
  \label{fig:5d-Veff}
\end{figure}

For \(r < 0\) (the broken phase), the tree-level potential takes a double-well shape.
Owing to the supremum definition of the Legendre transform, the non-convex region of the tree-level potential becomes constant in the effective potential \(V(\phi_c)\) in analogy with the classical Maxwell rule for coexisting phases~\cite{duncan2012conceptual,Orlando:2021usz}.
All in all, in the unbroken phase and at criticality (\emph{i.e.} for \(r \ge 0 \)) we have
\begin{equation}
  \Upsilon(x)=
  \begin{cases}
    \eval{ x y - \varpi (y)}_{y = y(x)} & \text{for \(x < \bar x\),} \\
    + \infty & \text{for \(x \ge \bar x\).}
  \end{cases}
\end{equation}
In the broken phase \(r < 0\), there are three distinct regions:
\begin{equation}
  \Upsilon(x)=
  \begin{cases}
    -\frac{N r^2}{2u} & \text{for \( 0 < x < - \frac{r N}{u} \),} \\
    \eval{ x y - \varpi (y)}_{y = y(x)} & \text{for \(- \frac{r N}{u} < x < \bar x\) ,} \\
    + \infty & \text{for \(x \ge \bar x\),}
  \end{cases}
\end{equation}
see Figure~\ref{fig:5d-Veff}.

For concreteness, we consider the case of \(D = 5\), where we have \(\Gamma(-5/2) = - 8 \sqrt{\pi}/15\), so that the grand potential is
\begin{equation}
   \omega(m) = (2N) \pqty{- \frac{m^5}{120 \pi^2} + \frac{(m^2 - r)^2}{4 u} }
\end{equation}
and there is a flex for
\begin{align}
  \bar y &= \frac{(4 \pi)^4}{u^2}, & \bar x &= \frac{ (4 \pi)^4}{3 u^3} N - \frac{r N}{u} .  
\end{align}
The effective potential describes a box with infinitely high walls of width
\begin{equation}
  2 \bar \phi_c =  2 \sqrt{\frac{N}{3}} \frac{(4 \pi)^2}{u^{3/2}} \sqrt{1 - \frac{3 r u^2}{(4\pi)^4} } ,
\end{equation}
signaling the breakdown of the \ac{eft} which requires a \ac{uv} completion.
The infinite walls hide the unphysical region corresponding to large values of \(u\) that contains the ostensible fixed point.

Within the box, the effective potential can be computed explicitly and expressed in terms of trigonometric functions:
\begin{equation}
  \begin{aligned}
    V(\phi_c) ={}& \Upsilon(\phi_c^2) = \sup_{y \in \setR^+}( \phi_c^2 y - \varpi (y))\\
    ={}&
    \begin{multlined}[t][\arraycolsep]
      \frac{2^{11}\pi^8 N}{5 u^5} \Big( 4 \sin(\frac{10 \theta + \pi}{6} ) + 20 \sin(\frac{4 \theta + \pi}{6} ) + 10 \cos(\frac{4 \theta + \pi}{3} ) + 20 \cos(\frac{\theta + \pi}{3} )\\
      - 20 \cos(\theta) - 7 \Big) - \frac{N r^2}{2u} ,
    \end{multlined}
  \end{aligned}
\end{equation}
where
\begin{equation}
  \cos(\theta) = 1 - 2 \frac{\phi_c^2}{\bar \phi_c^2} - \frac{6 r u^2 }{(4\pi)^4} \pqty{1 - \frac{\phi_c^2}{\bar \phi_c^2}} .  
\end{equation}
This result is much more transparent if we expand it in a series for small \(u\) and small \(r\), \emph{i.e.} around the free \ac{ir} fixed point.
The effective potential contains all the (infinitely many) leading-N corrections to the tree-level potential:
\begin{multline}
  V(\phi_c) = \left(\phi_c ^2 r+\frac{N r^{5/2}}{60 \pi ^2}\dots\right) + \left(\frac{\phi_c
   ^4}{2 N} + \frac{\phi_c ^2 r^{3/2}}{24 \pi ^2}+\frac{N r^3}{1152 \pi
   ^4}\dots\right) u \\
    + \left(\frac{\phi_c ^4 \sqrt{r}}{32 N \pi^2}
+\frac{\phi_c ^2 r^2}{384 \pi ^4}\dots\right) u^2 + \dots %
\end{multline}
In the critical phase (\(r = 0\)) we have an expansion in the only possible dimensionless combination \((u^{3/2} \phi_c)\):
\begin{equation}
  V(\phi_c) = \frac{u \phi_c ^4}{2 N} \pqty{ 1 + \frac{4}{5} \pqty{\frac{u^{3/2} \phi_c}{24 \pi^2 \sqrt{N} }} + \pqty{ \frac{u^{3/2} \phi_c}{24 \pi^2 \sqrt{N} } }^2 + \frac{3}{2} \pqty{ \frac{u^{3/2} \phi_c}{24 \pi^2 \sqrt{N} } }^3 + \frac{5}{2} \pqty{ \frac{u^{3/2} \phi_c}{24 \pi^2 \sqrt{N} } }^4   + \dots } .
\end{equation}
This expansion can also be interpreted in terms of Feynman diagrams around the free \ac{ir} fixed point.

Beyond \(\phi_c = \bar \phi_c\) (which vanishes in the \ac{uv} limit \(u \to \infty\)), the effective potential obtained with the supremum definition of the Legendre transform is literally infinite.
In other words, the theory needs a \ac{uv} completion.

We can however decide to extend the effective potential to a complex function and admit complex solutions for the maximization condition
\begin{equation}
  x = \varpi'(y) .
\end{equation}
Now this equation describes a Riemann surface and \(x = \bar x\) is a branch point that is joined to infinity by a branch cut.
For example, we can choose one of the branches and can expand the complex function \(V(\phi_c) = \Upsilon(\sqrt{\phi_c})\) around \(u \to \infty\), where the strongly coupled \ac{uv} \ac{cft} is expected to live, to find
\begin{multline}\label{eq:5DeffPotential}
  V(\phi_c) = \frac{12}{5} \pqty{\frac{3 \pi^2}{N}}^{2/3} e^{ 2  \pi i/3}  \phi ^{10/3}  \Bigg( 1 + 10 e^{ \pi i /3} \pqty{\frac{N \pi^4}{3} }^{1/3} \frac{1}{ \phi ^{2/3}
   u} \\ + 80 e^{2 \pi i /3} \pqty{\frac{N \pi^4}{3} }^{2/3} \frac{1}{ \phi ^{4/3}
   u^2} -\frac{1600 N \pi ^4}{9 \phi ^2
   u^3} + \order{u^{-4}} \Bigg) .
\end{multline}
In the next section we will use the leading term of this expansion to compute the conformal dimension of the lowest operator of given charge \(Q\) and show perfect agreement with the results in the literature.

\section{The strongly-coupled fixed point}
\label{sec:critical}

We have seen in Section~\ref{sec:EFT-Legendre-convexity} that the free energy density (at fixed charge density) can be obtained via a Legendre transform of the grand potential density \(\omega\) that we have used for the effective potential, but this time with respect to the variable \(m\).
Since at the conformal point we are interested in the scaling dimensions, we use the free energy instead of its density:
\begin{equation}
  F(Q) = \sup_{m > 0} ( Q m - \Omega(m)) ,
\end{equation}
where \(\Omega(m) = V_{\mathcal{M}} \omega(m)\).
Since the Legendre transform preserves homogeneity, the free energy as function of the charge $F = F(Q)$ has the same functional form as the free energy density as function of the charge density \(f=f(\rho)\) in Eq.~\eqref{eq:EnergyDensityAsLegendreTransform}. 
In a \ac{cft}, the physics on \(\setR^{d+1}\) is equivalent to the physics on the cylinder \(\setR \times S^d(r_0)\) and the free energy on the sphere is identified via the state-operator correspondence with the conformal dimension of the lowest operator of given charge \(Q\):\footnote{One can alternatively derive the conformal dimensions directly from the effective potential using the Callan--Symanzik equations, as it is done for example in~\cite{Abel:2017ujy}.}
\begin{equation}
  \Delta(Q) = r_0F_{S^d}(Q) .
\end{equation}

In general the convexity of \(\Omega\) as a function of \(m^2\) --- discussed in the previous section --- and the convexity of \(\Omega\) as a function of \(m\) are non-trivially related.
There might be regions in which \(\Omega(m)\) is not convex, however this \emph{does not signal a breakdown of the theory}.
The supremum definition of the Legendre transform always leads to a free energy that is convex as a function of a positive charge \(Q\) (or at worst equal to \(+\infty\)). It is important to note that the fixed-chemical potential regime described by \(\Omega(m)\) and the fixed-charge regime described by \(F(Q)\) are in general different.
Regions in which \(\Omega(m)\) is not convex cannot be reached by fixing the charge (see \emph{e.g.} the discussion in the appendix of~\cite{Gaume:2020bmp}).

Even if \(\Omega\) is convex, there can be regions that are not accessible at fixed charge.
The charge is positive and the maximization equation \(Q = \Omega'(m)\) has no solutions for values of \(m\) where \(\Omega(m)\) is decreasing.
We will see that this is the case for the model in \(D =3\) dimensions. The boundary of the accessible region (where \(\Omega'(m)>0\)) is at $m^2=1/(4r_0^2)$. This is precisely the value to which $r$ needs to be fine-tuned to flow to the conformal fixed point. 
More in general, $\Omega$ can also have concave regions even in a well-defined theory. An explicit example is the massive phase in \(D =3\) which is investigated in~\cite{Orlando:2021usz}.

The grand potential is again given in terms of a zeta function, this time for the Laplacian on the sphere:
\begin{equation}
  \Omega(m) = - N \zeta(-\sfrac{1}{2}| S^d, m) .
\end{equation}
For \(d\) even, there exist both a series expansion with a finite radius of convergence for small values of \(m\), and an asymptotic expansion for \(m \to \infty\).
We will separately discuss the cases \(d = 2\) and \(d = 4\) that, as we have seen in the previous section, correspond respectively to a well-defined and a non-unitary theory.

\subsection{Three dimensions: the zeta function on $S^2$}
\label{sec:three-dimens-zeta}

\begin{figure}
  \centering
  \includestandalone[mode=image]{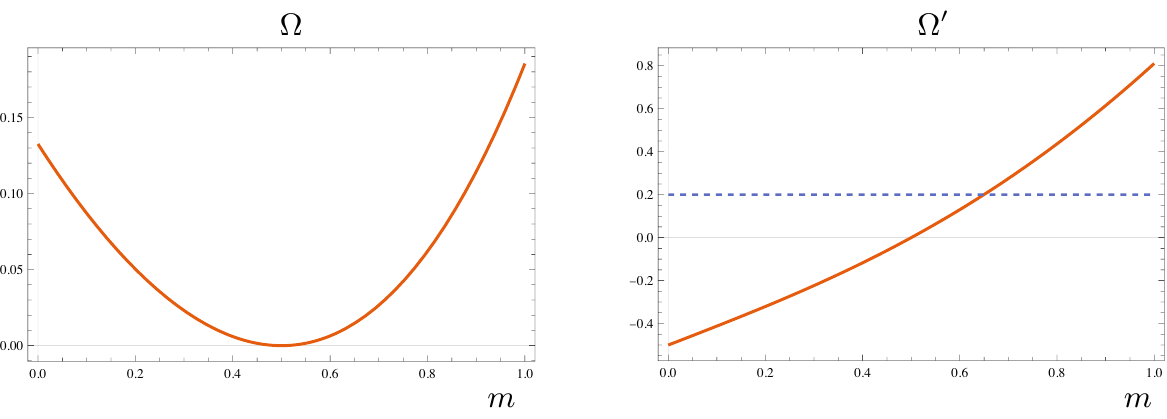}
  \caption{The grand potential \(\Omega(m)\) (left) and its first derivative \(\Omega'(m)\) (right) on the two-sphere.
    The function \(\Omega(m)\) is convex for all values of \(m\) and has a minimum for \(m = 1/(2r_0)\).
    The maximization condition has a solution with \(Q \ge 0\) in the region \(m \ge 1/(2r_0) \) (right). }
  \label{fig:S2-omega}
\end{figure}

The zeta function for the two-sphere has been discussed in~\cite{Alvarez-Gaume:2019biu,Dondi:2021buw}.
It cannot be expressed in terms of elementary functions, but we can give an asymptotic expansion for large values of \(m\) and a convergent expansion at small \(m\):
\begin{itemize}
\item At large \(m\) we write \(\zeta(s \mid S^2, m)\) as a Mellin transform of the heat kernel:
\begin{equation}
  \zeta(s \mid  S^2, m) = \frac{1}{\Gamma(s)} \int_0^\infty \frac{\dd{t}}{t} t^s e^{-m^2 t} \Tr[e^{t \Laplacian_{S^2}{} }] .
\end{equation}
Using the known form of the asymptotic expansion for small \(t\) of the heat kernel,
\begin{equation}
  g(t) = 
  \Tr[ e^{t \pqty{\Laplacian{} - \sfrac{1}{4r_0^2}}}] \sim \frac{r_0^2}{t} - \sum_{n=1}^\infty \frac{(-1)^n(1-2^{1-2n})}{n! r_0^{2n-2}} B_{2n} t^{n-1} \equiv \frac{r_0^2}{t}  \sum_{n=0}^\infty a_n \pqty{\frac{t}{r_0^2}}^n ,
\end{equation}
we derive the large-\(m\) asymptotic expansion for the grand potential
\begin{equation}
  \label{eq:large-m-grand-potential}
    \Omega(m) = - N \zeta ( - \sfrac{1}{2} \mid S^2 , m^2 ) = \frac{2N}{r_0}  \pqty{m^2 r_0^2 - \frac{1}{4} }^{3/2} \sum_{n=0}^\infty \frac{ \Omega_n}{(m^2 r_0^2 - 1/4)^{n}} ,
\end{equation}
where the coefficients are given by
\begin{equation}
  \Omega_n =- \frac{1}{4 \pi}  \sum_{k \neq 0} \frac{(-1)^k}{(k\pi)^{2n}} \Gamma\left( n + \frac{1}{2} \right) \Gamma \left( n - \frac{3}{2} \right) .
  \label{eq:gran-pot-coeff}
\end{equation}
\item In the opposite limit of small \(m\), one can instead derive a convergent expansion from the definition in terms of eigenvalues:
\begin{multline}
  \Omega(m) = - N \zeta(-\sfrac{1}{2}| S^2, m) = - N \eval{ \sum_{l=0}^\infty (2l + 1) \left( \frac{l(l+1)}{r_0^2} + m^2 \right)^{-s}}_{s = -1/2} \\
  = - 2N  r_0^{2s} \eval{ \sum_{k=0}^\infty \binom{-s}{k} \zeta(2s+2k-1; \sfrac{1}{2}) \pqty{m^2 r_0^2 - \frac{1}{4}}^k}_{s=-1/2} ,
\end{multline}
where \(\zeta(s ; a)\) is the Hurwitz zeta function
\begin{equation}
  \zeta(s ; a) = \sum_{n=0}^\infty ( n + a)^{-s} .
\end{equation}
\end{itemize}
The explicit form of the convergent expansion shows that \(\Omega(m)\) is always convex and has a minimum for \(m = 1/(2r_0)\), where it also vanishes:
\begin{align}
  \eval{\Omega(m)}_{m = 1/(2r_0)} &= 0, & \eval{\Omega'(m)}_{m = 1/(2r_0)} &= 0 .
\end{align}
We can perform the Legendre transform order by order and find a solution to the maximization condition \(Q = \Omega'(m)\) (i.e. \(\rho = \omega'(m)\)) for positive values of \(Q\) in the region \(m > 1/(2 r_0)\) (see Figure~\ref{fig:S2-omega}).
The boundary \(m^2 = 1/(4r_0^2)\) is precisely the value of the conformal coupling \(m^2 = R/8\) to which $r$ needs to be fine-tuned in the \ac{uv} action for the model to flow to the \ac{wf} fixed point in the \ac{ir}.

The explicit expressions for \(F(Q)\) in the asymptotic \(Q \gg 1\) and in the convergent \(Q \ll 1\) regions have been computed in~\cite{Alvarez-Gaume:2019biu}:
\begin{align}
  r_0 \frac{F(Q)}{2N} &= \frac{2}{3} \pqty{\frac{Q}{2N} }^{3/2} + \frac{1}{6} \pqty{\frac{Q}{2N} }^{1/2} - \frac{7}{720 } \pqty{\frac{Q}{2N} }^{-1/2} + \dots \\
  r_0 \frac{F(Q)}{2N}  &= \frac{1}{2} \pqty{\frac{Q}{2N}}  + \frac{4}{\pi ^2} \pqty{\frac{Q}{2N} }^2 + \frac{16 \left(\pi ^2-12\right) }{3 \pi ^4} \pqty{\frac{Q}{2N} }^3 + \dots
\end{align}
A resurgence analysis has been performed in~\cite{Dondi:2021buw} to interpolate between the two regions.

\subsection{Five dimensions: the zeta function on $S^4$}
\label{sec:five}

For \(d = 4\) we have seen in the previous section that the Legendre transform leads to an inaccessible region that hides the expected non-trivial \ac{uv} fixed point.
One can however continue the effective potential \(V(\phi_c)\) analytically and study the behavior of the theory around \(u \to \infty\) starting from the zeta function on \(S^4\) in analogy to the three-dimensional case above.
\subsubsection{Expansions of the zeta function}
\label{sec:zeta-S4}

The heat kernel traces for even-dimensional spheres are related by a recursion relation~\cite{Cahn1975} which we can use starting from the results of the two-sphere:
\begin{equation}
  \label{eq:HeatKernelForEvenDimSpheresGeneral}
  \begin{aligned}
    \Tr[ e^{\Laplacian_{S^{2n}} t} ] (t) &= \sum_{l\geq 0} \left[ \frac{2l+2n-1}{2n-1} \prod_{k=1}^{2n-2} \frac{l+k}{k} \right]  e^{- l(l+2n-1) t/{r_0^2}}  \\
    &= \frac{e^{\left( n - \sfrac{1}{2} \right)^2 t/r_0^2}}{(2n-1)!}   \sum_{j=0}^{n-1} \beta_{j;n}   (-1)^j  r_0^{2j} \dv[j]{g(t)}{t} ,
  \end{aligned}
\end{equation}
where \(g(t)\) is the trace of the heat kernel that we have seen above,
\begin{equation}
  g(t) = \Tr[ e^{ t \pqty{\Laplacian_{S^2}{} - \sfrac{1}{4 r_0^2} }} ]  ,
\end{equation}
and the coefficients $\beta_{j;n}$ are defined by the relationship
\begin{equation}
\frac{2s}{(2n-1)! } \prod_{j=\frac{1}{2} , \frac{3}{2},\dots}^{n-3/2} \left( s^2 - j^2\right) = \frac{2s}{(2n-1)! } \sum_{j=0}^{n-1} \beta_{j;n}   s^{2j} .
\end{equation}
From a generalization of this relation one can derive a similar formula for the zeta function on a generic sphere:\footnote{This expression relies on an integration by parts and needs to be suitably continued for $m^2r_0^2 < \pqty{n-\sfrac{1}{2} }^2$.}
\begin{equation}
\begin{aligned}
  \zeta(s \mid S^{2n},m) = \frac{r_0^{2s}}{(2n-1)!} \sum_{j=0}^{n-1} \beta_{j;n}  \sum_{k=0}^j & \binom{j}{k} (-1)^k \pqty{ m^2r_0^2 - \pqty{n-\sfrac{1}{2} }^2}^{j-k} \\
  &  \times r_0^{2k} \zeta\left(s-k \;\middle|\; S^2, \sqrt{m^2 - \pqty{n^2 -n} r_0^{-2} } \right) .
\end{aligned}
\end{equation}
Once more there are two interesting regimes:

\paragraph{Large $m$.}
For large values of $m$ (which corresponds to large values of \(Q/N\)) we can  expand the trace of the heat kernel in the small $t$-limit, since the Mellin integral localizes around $t=0$.
On the four-sphere the relation Eq.~\eqref{eq:HeatKernelForEvenDimSpheresGeneral} gives ($\beta_{0;2}= -\frac{1}{4}$, $\beta_{1;2}= 1$)
\begin{equation}\label{eq:RelationshipFourSphereToTwoSphere}
\Tr[ e^{\Laplacian_{S^{4}} t} ] =  \frac{e^{\sfrac{9}{4r_0^2} t}}{6} \left( \beta_{0;2} - \beta_{1;2} r_0^2 \dv{t} \right) g(t) .
 \end{equation}
The zeta function in the Mellin representation is then
\begin{equation} \label{eq:FullAsymptoticExpansionFourSphere}
  \begin{aligned}
\zeta(s   \mid  S^4 , m ) ={}& -\frac{1}{\Gamma(s)} \int_0^\infty \frac{\dd{t} }{t}   t^s e^{-m^2 t} \pqty{ \frac{e^{\sfrac{9}{4r_0^2} t}}{24} g(t) + \frac{e^{\sfrac{9}{4r_0^2} t}}{6} r_0^2 \dv{g(t)}{t} } \\
  ={} & \frac{r_0^{2s}}{6(s-1)(s-2)} \Big( m^2r_0^2 - \frac{9}{4} \Big)^{2-s} \\
  &+ \frac{ r_0^{2s}}{24}  \sum_{k\geq 0} \frac{(-1)^{k}   \left( 2^{2k+1} -1 \right) B_{2k}}{ 2^{2k-1}  \left( m^2r_0^2 - \frac{9}{4} \right)^{k+s-1} } \frac{\Gamma(s+k-1)}{\Gamma(s) k!}  \bigg[  \frac{\left( 2^{2k-1} -1 \right)}{\left( 2^{2k+1} -1 \right)}  - \frac{k B_{2k+2}}{ \pqty{k+1} B_{2k}}\bigg] .
  \end{aligned}
\end{equation}
This expansion is asymptotic and can be studied using resurgence, see Appendix~\ref{sec:resurgence}.

\paragraph{Small $m$.}
The other interesting limit is for small values of $m$, where the zeta function is expanded into a convergent series valid in the regime $0< m r_0 < 3/\sqrt{2}$:
\begin{equation}
  \begin{aligned}
    \zeta (s \mid S^{4} , m ) &= \sum_{l\geq 0} \left[ \frac{2l+3}{3} \prod_{k=1}^{2} \frac{l+k}{k} \right]  \left( \frac{ l(l+3)}{r_0^2}  + m^2 \right)^{-s} \\
    &=\frac{r_0^{2s}}{3} \sum_{k\geq0}  {-s \choose k}   \left[ \zeta \left(2s +2k -3 ; \sfrac{3}{2} \right) - \frac14  \zeta \left(2s +2k -1 ; \sfrac{3}{2} \right) \right]  \left(m^2 r_0^2 - \frac{9}{4} \right)^{k} .
  \end{aligned}
\end{equation}

The two limits can be understood as expansions of a single function that can be obtained using resurgence techniques, as we show in Appendix~\ref{sec:resurgence}.
This function can be written in terms of the principal value of an integral involving Bessel's functions.
The grand potential for example has the form
\begin{equation}
  \frac{\Omega(m)}{2N} = - \frac{m^4 r_0^3}{24\pi}  \text{P.V.}\underset{0}{\overset{\infty}{\int}} \frac{ \dd{y} }{y  \sin(y) } \bigg( 2 K_4 (2m r_0 y)  + \pqty{2 +\frac{1}{m^2 r_0^2}} K_2 (2m r_0 y) \bigg). 
\end{equation}

\begin{figure}
  \centering
  \includestandalone[mode=image]{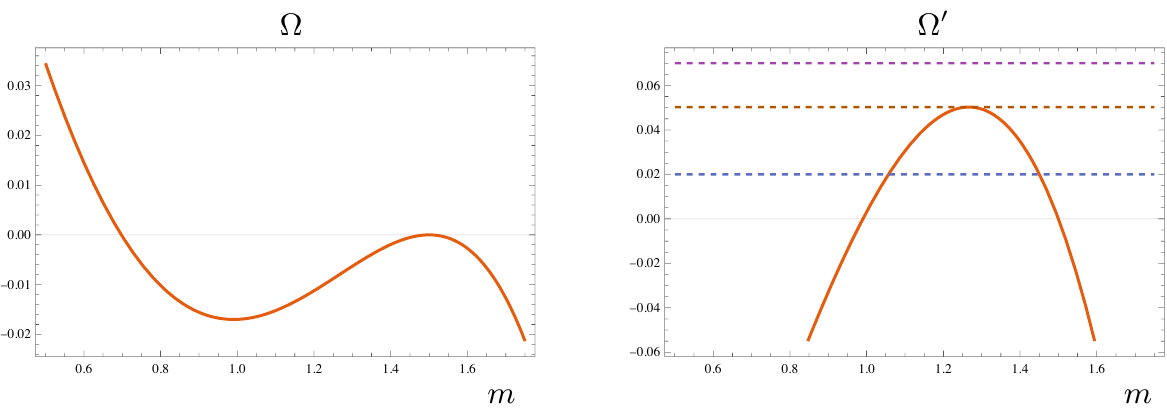}
  \caption{The grand potential \(\Omega(m)\) (left) and its first derivative \(\Omega'(m)\) (right) on the four-sphere.
    The function \(\Omega(m)\) is convex for \(m \le m_{\text{fl}} \approx 1.266\dots\), has a minimum for \(mr_0 = m_{\text{min}}r_0 \approx 0.9927 \) and a maximum for \(m r_0 = m_{\text{max}}r_0 = 3/2\).
    The maximization condition admits two solutions for \(0 \le Q < Q_{\text{fl}}  = \Omega'(m_{\text{fl}}) \approx (2N)\cdot 0.5029\dots  \) and no solutions beyond \(Q = Q_{\text{fl}}\) (right).
    The leftmost solution gives a convex function, but does not satisfy \(F(0) = 0\) since \(F(0) = -\Omega(m_{\text{min}} ) \approx 0.017\dots > 0\).
  }
  \label{fig:S4-omega}
\end{figure}

\subsubsection{Free energy and conformal dimensions}
\label{sec:F-Delta-S4}

One interesting observation is that in the regime where the small-\(m\) expansion is convergent and can be trusted, the zeta function on the four-sphere has a more interesting behavior than the two-sphere case: there is a maximum and a minimum, \emph{i.e.} a convex and a concave region separated by a flex (see Figure~\ref{fig:S4-omega}).
This happens for the values
\begin{align}
  r_0 m_{\text{min}} &\approx 0.9927\dots &  r_0 m_{\text{fl}} &\approx 1.266\dots & r_0 m_{\text{max}} &= \frac{3}{2} .
\end{align}
The value of the flex can also be found with quite good precision using an optimal truncation of the large-\(m\) expansion of the grand potential (see Appendix~\ref{sec:resurgence}).
Note however that this expansion cannot reproduce the whole structure of \(\Omega(m)\) and in particular it cannot reproduce the maximum at \(r_0 m_{\text{max}} = 3/2 \).

By an argument that is by now familiar, the Legendre transform is non-trivial only for values of \(Q\) smaller than
\begin{equation}
  \frac{Q_{\text{fl}}}{2N} = \Omega'(m_{\text{fl}}) \approx 0.05029\dots
\end{equation}
The free energy is equal to \(+\infty\) for \(Q > Q_{\text{fl}}\) since in this case, \(\Omega(m)\) is not bounded below:
\begin{equation}
  \frac{F (Q)}{2N} =
  \begin{cases}
    \eval{Q m - \Omega(m)}_{m = m(Q)} & \text{for \(0 < Q < Q_{\text{fl}} \approx (2N)\cdot 0.05029\dots \)} \\
    + \infty & \text{for \(Q > Q_{\text{fl}}\).}
  \end{cases}
\end{equation}
However, the function \(F(Q)/(2N)\) that we obtain in this way cannot be describing the conformal dimension of operators of fixed charge in a \ac{cft}.
By the state-operator correspondence, \(r_0 F(0)\) is the dimension of the identity operator and has to vanish, while we have
\begin{equation}
  \frac{r_0 F(0)}{2N} = - \frac{1}{2N} \Omega(m_{\text{min}}) \approx 0.01699\dots{} \neq 0 .
\end{equation}
The appearance of the constant term means that we are not at the critical point.
Expanding in $Q$, we find
\begin{equation}
  \frac{r_0 F(Q)}{2N} \approx 0.01699\dots{} + 0.9903\dots \frac{Q}{2N}  + 1.516\dots \pqty{\frac{Q}{2N} }^2 + \dots 
\end{equation}
We see that the leading term in $Q$ is linear, which means that we are in a massive phase.

Alternatively, we could decide to ignore convexity and use the naive definition of Legendre transform and expand around the maximum \(m_{\text{max}}^2 = 9/(4 r_0^2)\), which is precisely the value of the conformal coupling and where \(\Omega(m_{\text{max}}) = 0\).
In this case we would find
\begin{equation}
  r_0 F(Q) = 2N \left[  \frac{3}{2} \left( \frac{Q}{2N} \right)  - \frac{32}{3 \pi^2} \left(\frac{Q}{2N}\right)^2 + \order{Q^3} \right] .
\end{equation}
This same result has been found in~\cite{Giombi:2020enj} (where \(Q_{\text{fl}}\) has been identified as the critical value above which \(\Delta(Q)\) becomes complex), and it is consistent with the conformal dimension of the \(Q\)-th power of a field of dimension \(3/2\) (see also~\cite{Arias-Tamargo:2020fow}).
It would be interesting to understand better the meaning of this expression.
On the one hand we are discussing a \ac{uv} continuation obtained via a hard-to-justify complex effective potential;
on the other hand we are expanding around a maximum, in a concave region of \(\Omega(m)\) that one would not expect to contribute to the Legendre transform.
Once more we can try and continue \(F(Q)\) to a complex function and look at \(Q_{\text{fl}}\) as the beginning of a branch cut.
In this way we can study the behavior for large values of \(Q\) using the large-\(m\) expansion in each of the four branches.
Using the expansion in Eq.~(\ref{eq:FullAsymptoticExpansionFourSphere}) we find
\begin{equation}
\Delta(Q) = r_0 F_{S^4} (Q) =   2N  \left[  f_1 \frac{4\sqrt{3} }{5}  \left( \frac{Q}{2N} \right)^{\frac{5}{4}}  - \frac{f_2}{\sqrt{3}} \left( \frac{Q}{2N} \right)^{\frac{3}{4}} \right] ,
\end{equation}
where the phases \(f_1\) and \(f_2\) depend on the choice of the branches:
\begin{equation}
\label{eq:branch-table}
  \begin{tabular}{llllll}
    &branch 1 & branch 2 & branch 3 & branch 4 \\
    \(f_1\) &\(e^{i \pi/4}\)&\(e^{-i \pi/4}\)&\(e^{3\pi i/4}\)& \(e^{-3\pi i/4}\) \\
    \(f_2\) &\(e^{3i \pi/4}\)& \(e^{-3i \pi/4}\)&\(e^{\pi i/4}\)& \(e^{-\pi i/4}\)
  \end{tabular}
\end{equation}
which again agrees with the results in~\cite{Giombi:2020enj}. 
The branch choice here must correspond to the one that we have already made in Eq.~\eqref{eq:5DeffPotential} since the coefficient of the \(\phi^{10/3}\) term in the effective potential completely fixes the coefficient of the \(Q^{5/4}\) term in the conformal dimension.
To see that, observe that at criticality, for dimensional reasons, the effective potential must have the form
\begin{equation}
	V(\phi_c ) = \kappa \phi^{10/3} \ ,
\end{equation}
which corresponds to a fixed-charge vacuum energy on the torus
\begin{equation}
	E_{T^4} = \frac{2^{7/4} \sqrt{3} \kappa^{3/8}}{5 N^{1/4} L} Q^{5/4} \ .
\end{equation}
But since the ground state is homogeneous, this implies that the leading term in the large-charge expansion of the energy on the four-sphere is
\begin{equation}
	\eval{E_{S^4}}_{Q^{5/4}} = \frac{L}{V_{S^4}^{1/4}} E_{T^4} = \frac{2 \kappa^{3/8}}{3^{1/8} 5^{5/8} \sqrt{\pi} r_0} = \eval{\frac{\Delta(Q)}{r_0}}_{Q^{5/4}}.
\end{equation}
In particular, the special choice we have made in Eq.~\eqref{eq:5DeffPotential} corresponds to the first branch in Eq.~\eqref{eq:branch-table}.

\subsubsection{A phase transition on the cylinder}

Further insights on the nature of the critical point can be obtained moving away from the limit \(u \to \infty \).
One possible interpretation for a complex effective potential (obtained via the naive Legendre transform) is in terms of unstable states (see \emph{e.g.}~\cite{Coleman:1988aos}).
If we follow a given state while changing a parameter in the theory, it can be that at some point the state becomes unstable and its energy develops and imaginary part that can be understood in terms of decay rates.
Something similar happens here, were states that are perfectly fine at small \(u\) become unstable in the \ac{uv}.
Having an explicit (convergent) expansion for the zeta function on the sphere at small \(m\) we can see how this happens and the implications for the free energy on the sphere at fixed (small) charge \(Q\).\\
In general, if a function \(\Omega(m)\) has a critical point for \(m = m_c\), its Legendre transform \(F(Q)\) can be expanded around \(Q = 0\) in terms of local properties of \(\Omega\):
\begin{equation}
  F(Q) =  \eval{\bqty{ - \Omega(m_c) + m_c Q + \frac{1}{2 \Omega''(m_c)} Q^2 + \dots }}_{m_c = 3/(2r_0)} .
\end{equation}
Keeping only the first two terms, this is the free energy of a fixed-charge state for a system describing a particle of mass \(m_c\).

\begin{figure}
  \centering
  \includestandalone[mode=image]{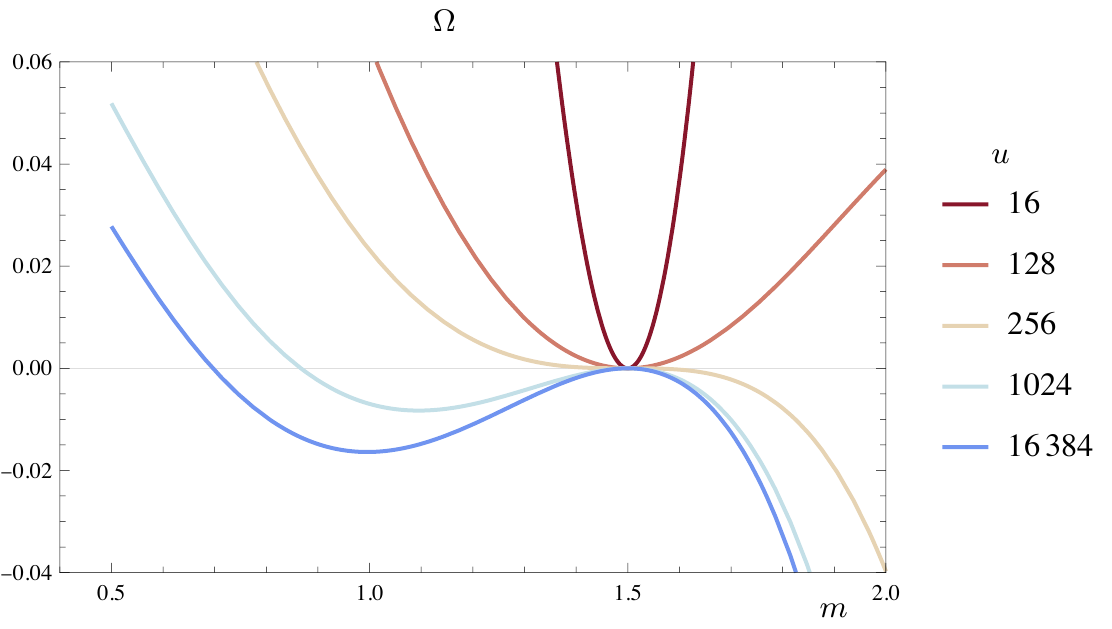}
  \caption{The grand potential \(\Omega\) on the cylinder $\mathbbm{R}\times S^4_{r_0}$ for \(r = 9/(4r_0^2)\) at different values of \(u\). For \(u < 256 r_0\), the point \(m = 3/(2r_0)\) is a local minimum. For larger values of \(u\) a new minimum appears for \(m < 3/(2 r_0)\). }
  \label{fig:grand-potential-S4-u}
\end{figure}

If we fine tune \(r = 9/(4r_0^2)\), which is the conformal coupling on the sphere, the grand potential takes the form
\begin{equation}
  \Omega(m) = (2N) \pqty{ - \frac{1}{2} \zeta(- \sfrac{1}{2}| S^4, m) + V_{S^4} \frac{(m^2 - 9/(4r_0^2))^2}{4 u} } .
\end{equation}
The zeta function and its first derivative vanish at \(m^2 =9/(4r_0^2)\) and around that point we have the perturbative expansion:
\begin{equation}
  \Omega(m) = \frac{2N}{r_0} \bqty{ \frac{\pi^2}{384} \pqty{\frac{256 r_0}{u} - 1 } \pqty{m^2 r_0^2 - \frac{9}{4} }^2 + \frac{(\pi^2 - 12) \pi^2}{2^8 3^2} \pqty{m^2 r_0^2 - \frac{9}{4} }^3 + \dots } .
\end{equation}
The nature of the critical point \(m_c = 3/(2r_0)\) changes with the value of \(u\).
For \(u < 256 r_0\) we have a local minimum, which turns into a local maximum for larger values of the coupling where a new local minimum appears for \(m < 3/(2r_0)\) (see Figure~\ref{fig:grand-potential-S4-u}).

The supremum definition of the Legendre transform tracks the position of the minimum: for small values of \(u\) the free energy is determined by the behavior around \(m = 3/(2 r_0)\),
\begin{equation}
  r_0 F(Q) = \frac{3}{2} Q + \frac{192 u}{\pi^2 \pqty{256 r_0 - u}}  \frac{Q^2}{2N} + \dots
\end{equation}
as long as the coefficient of \(Q^2\) is positive.
For large \(u\), the free energy depends on the local properties of the grand potential around the new minimum appearing at \(m < 3/(2r_0)\) which, as we have seen,  describes a massive phase in the \(u \to \infty \) limit.
However, this new minimum is metastable and, for large enough charge, the free energy becomes infinite since the grand potential is not bounded below.

Using  the naive definition of Legendre transform, on the other hand, for large \(u\) we can choose between two branches, one of which is still centered around \(m = 3/(2r_0)\) and corresponds to a concave (unstable) region of the grand potential.
For small values of \(Q\), the free energy is still real, but the coefficient of \(Q^2\) is negative, which is a sign of instability.
When expanding around a maximum, generically the free energy is concave as function of the charge. 
It seems that the absence of convexity is a general feature of an expansion around an unstable state, consistently with the conjecture of~\cite{Aharony:2021mpc}.

\subsection{$D = 4+ \epsilon$ and $D = 6 - \epsilon$}
\label{sec:epsilon-dimensions}

In Section~\ref{sec:4to6} we have seen that for \(4 < D < 6\) the expected \ac{uv} fixed point cannot be reached unless we continue the effective potential to a complex function with a branch cut.
Just like we have done in the case of \(D = 5\), we can still compute the free energy on the sphere, that we naturally identify with the conformal dimension of the lowest operator of given charge:
\begin{align}
  \Delta(Q) &= r_0 \Omega^*(Q) ,  & \Omega(m) &= - N \zeta(-\sfrac{1}{2} |S^{D-1} , m ) .
\end{align}
In the limit of large charge, the zeta function can be computed perturbatively in terms of geometrical invariants for any real value of \(D\) using Weyl's asymptotic expansion for a manifold of dimension \(d = D - 1\))~\cite{weyl1911asymptotische,Elizalde:2012zza,kirsten2001spectral,Vassilevich:2003xt}:
\begin{equation}
\Tr \left[ e^{ \Laplacian_{\mathcal{M}} t} \right] = \sum_{n=0, \frac{1}{2},1,\frac{3}{2},\dots} K^{\mathcal{M}}_n  t^{n -\frac{d}{2} } ,
\end{equation}
where the $K^{\mathcal{M}}_n$ are the heat kernel coefficients.
These coefficients can be expressed in terms of the geometry of the manifold. For manifolds without boundaries the half-integer coefficients $K^{\mathcal{M}}_{k/2}$ vanish.
The first two integer coefficients can be written in closed form as
\begin{align}
K^{\mathcal{M}}_0 &= \frac{V_{\mathcal{M}}}{(4\pi)^{d/2}}  , & K^{\mathcal{M}}_1 &= \frac{1}{6} \frac{1}{(4\pi)^{d/2}} \int_{\mathcal{M}} \dd{x}  R .
\end{align}
The leading term in the asymptotic expansion of the zeta function on the $d$-sphere then reads
\begin{equation}
\zeta(s\mid  S^d , m ) = \sum_{n\geq0} K^{S^d}_n \frac{\Gamma(s+n -\frac{d}{2} ) }{\Gamma(s)} m^{d-2n-2s  }  = K^{S^d}_0  \frac{\Gamma(s-d/2) }{\Gamma(s)} m^{d-2s}  + \dots 
\end{equation}

As we have observed in Section~\ref{sec:phi4}, for $s=\pm 1/2$ and $d$ odd, the Gamma function in the numerator diverges.
We can however resort to analytical continuation in \(d\) and study the behavior around \(D =4 \) and \(D=6\), \emph{i.e.} we expand the gamma function around the spatial dimensions for $d = 3 \pm \epsilon$ and $d = 5 \pm \epsilon$.
The geometrical invariants are
\begin{align}
  K^{S^{3\pm\epsilon}}_0 m^{3\pm \epsilon} &=  \frac{2^{-2\mp\epsilon} \sqrt{\pi}}{\Gamma (2\pm\frac{\epsilon}{2})} (r_0 m)^{3\pm\epsilon} = \frac{\sqrt{\pi}}{4}\left[ 1 \pm \left( \frac{\gamma_{_{\text{EM}}}}{2} - \log(2) - \frac{1}{2} + \log(r_0m) \right) \epsilon \right] (r_0m)^{3} ,  \\
  K^{S^{5\pm\epsilon}}_0 m^{5\pm \epsilon} &= \frac{2^{-4 \mp \epsilon} \sqrt{\pi}}{\Gamma (3\pm\frac{\epsilon}{2})}(r_0 m)^{5\pm\epsilon} = \frac{\sqrt{\pi}}{32} \left[ 1 \mp \left(\log(2) + \frac{3}{4} - \frac{\gamma_{_{\text{EM}}}}{2} - \log(r_0m) \right) \epsilon \right] (r_0m)^{5} .
\end{align}
Using the fact that $\Gamma(\epsilon-n) = (-1)^n \epsilon^{-1}/n! + \order{\epsilon^{0}}$, the first order result for the zeta functions we are looking for is
\begin{align}
  \zeta\left(\sfrac{1}{2} \mid S^{3\pm\epsilon} , m \right) &= \pm \epsilon^{-1} \frac{ r_0^3 m^2 }{2} + \dots , & \zeta\left(-\sfrac{1}{2} \mid S^{3\pm\epsilon} , m \right) &= \pm \epsilon^{-1} \frac{ r_0^3 m^4 }{8} + \dots ,  \\
  \zeta\left(\sfrac{1}{2} \mid S^{5\pm\epsilon} , m \right) &= \mp \epsilon^{-1} \frac{ r_0^5 m^4 }{32} + \dots , & \zeta\left(-\sfrac{1}{2}  | S^{5\pm\epsilon} , m \right) &= \mp \epsilon^{-1} \frac{ r_0^5 m^6 }{192} + \dots .
\end{align}

Around \(D = 4 \), the minimization condition gives \( Q/N = \mp (r_0 m)^3/(2 \epsilon)\), and, as expected, there is a solution for positive charge only in \( D < 4\).
The corresponding conformal dimension of the lowest operator of charge \(Q\) is
\begin{equation}
 \begin{cases}
   \Delta_{-\epsilon} (Q) =  \frac{3N}{2^{1/3}} \epsilon^{1/3}  \left( \frac{Q}{2N} \right)^{4/3}  + \dots & \text{for \(D  = 4 - \epsilon\)}, \\
   \Delta_{+\epsilon} (Q) =  \frac{3N}{2^{1/3}} e^{i \pi (2k + 1)/3}  \epsilon^{1/3}  \left( \frac{Q}{2N} \right)^{4/3}  + \dots& \text{for \( D = 4 + \epsilon\)},
\end{cases} 
\end{equation}
where \(k = 0, 1, 2\) depends on the choice of one of the three possible branches.

Similarly, around \(D = 6\), the minimization gives \(Q/N = \pm (r_0 m)^5/(32 \epsilon) \), where now the positive-charge solution exists for \(D > 6\).
The corresponding conformal dimension of the lowest operator is
\begin{equation}
  \begin{cases}
    \Delta_{-\epsilon} (Q) =  \frac{2^{6/5} 5 N}{3}  e^{i \pi (2k + 1)/5}  \epsilon^{1/5}  \left( \frac{Q}{2N} \right)^{6/5}  + \dots  & \text{for \(D  = 6- \epsilon\)}, \\
    \Delta_{+\epsilon} (Q) =  \frac{2^{6/5} 5 N}{3} \epsilon^{1/5} \left( \frac{Q}{2N} \right)^{6/5}  + \dots  & \text{for \(D  = 6 + \epsilon\)}, \\
\end{cases} 
\end{equation}
where now \(k = 0, 1, 2, 4, 5\) corresponds to the five possible branches.

These results agree with those in~\cite{Antipin:2020abu,Giombi:2020enj}.
As expected from the general argument of Section~\ref{sec:phi4}, for  $4<D<6$, the Legendre transform requires an analytic continuation and  leads to  complex conformal dimensions.

\section{Conclusions}
\label{sec:Conclusions}

In this paper we have discussed the phase diagram of the large-N $\varphi^4$ vector model in $2<D<6$.
Since we are working in the double-scaling limit $N\to \infty$, $Q\to \infty$, $Q/N$ fixed, we are able to do this without resorting to diagrammatic techniques and obtain a closed-form result which resums all the contributions at leading order in $N$. 
Having calculated the grand potential $\Omega$, we find the effective potential, which is independent of the charge fixing, via a Legendre transform. 
The convexity properties of $\Omega$ are related to the unitarity of the theory: in a unitary theory, the effective potential is always convex and the Legendre transform preserves convexity. 
For $2<D<4$, $\Omega$ is convex as a function of $m^2$, while for $4<D<6$ it has a flex and is not bounded from below.
This is consistent with earlier observations that in this range of dimensions the theory is non-unitary in the \ac{uv} and flows to a complex \ac{cft}.
Since we have access to the phase diagram, we can also study the free energy on the cylinder.
In $D=5$ we find a new metastable massive phase that appears at high energy.

\bigskip
There is a number of ways in which our observations can be extended:
\begin{itemize}
	\item The metastable phase and instabilities in $4<D<6$ could be explored in terms of instantons on the cylinder, extending the treatment in~\cite{Giombi:2019upv} beyond the critical point.
	\item Our construction is formally valid in any dimension. In $0<D<2$ the function $\Omega$ is convex, but it is unclear to us what is the fate of the theory in the \ac{ir}.
	\item Similarly, for $6<D<8$, $\Omega$ is again convex, but generically we do not expect a unitary theory since the collective Stratonovich field violates the unitarity bound. 
More in general, the convexity of $\Omega$ is satisfied for $4n-2<D<4n$.
	\item In this work, we have restricted ourselves to canonical kinetic terms when calculating the effective potential. It would be interesting to see if one can make statements also for general kinetic terms.
	\item All the physics is already encoded in the grand potential. If we could extract all the information meaningfully from $\Omega$, it may be possible to avoid calculating the effective potential altogether. 
\end{itemize}
We leave these points for future investigation.

\subsection*{Acknowledgements}

\begin{flushleft}
  \dosserif
  We would like to thank Luis~Álvarez-Gaumé, Nicola~Dondi, Simeon~Hellerman, Igor~Pesando and Uwe-Jens~Wiese for enlightening discussions and Tim~Schmidt for collaboration on a related project. We would also like to thank the referee for their comments that have helped improving this paper.
  The work of S.R. and R.M. is supported by the Swiss National Science Foundation under grant number 200021 192137.
  D.O. acknowledges partial support by the \textsc{nccr 51nf40--141869} ``The Mathematics of Physics'' (Swiss\textsc{map}).
\end{flushleft}

\appendix
\section{Resurgence of the four-sphere}
\label{sec:resurgence}

\paragraph{Borel Resummation.}
Borel resummation is a tool to transform an asymptotic expansion into a resurgent function (see~\cite{Dorigoni:2014hea} for an introduction).
Suppose we have a factorially divergent asymptotic series
\begin{align}\label{eq:LargeOrderBehaviorOfTheCoefficientsGeneral}
&\Phi_0(t) = \sum_n a_{n} t^{n} , \quad & a_n \sim \sum_k \frac{S_k}{2\pi i} \frac{\beta_k}{A_k^{n\beta_k + b_k} }  \sum_{l\geq 0} a_{l;k} A_k^l  \Gamma\big( n \beta_k + b_k - l \big) .
\end{align}
The assumption is that the asymptotic expansion can be completed into a trans-series of the form
\begin{align}\label{eq:TransSeriesExpansionGeneralForm}
&\Phi(S_k, t) = \Phi_0 (t) + \sum_{k\neq 0} \sigma_k  e^{-\frac{A_k}{ t^{1/\beta_k} }} t^{-b_k/\beta_k} \Phi_k (t) , &\Phi_k (t) \sim \sum_{l\geq0} a_{l;k} t^{l/\beta_k} ,
\end{align}
where the $\sigma_k$ are ambiguities that arise from the fact that for any value of these parameters the trans-series corresponds to the same asymptotic series $\Phi_0 (t)$.
The closed-form Borel transform of a series is defined by
\begin{equation}
\mathcal{B}[\Phi_0] (y) = \sum_{n\geq0} \frac{a_n}{\Gamma\big(\max_k(b_k) \cdot n + \max_k(\beta_k) \big)} y^n ,
\end{equation}
and the (directional) Borel resummation of the series $\Phi_0$ is defined as
\begin{equation}\label{eq:BorelTransformGeneralDefinition}
\mathcal{S}_{\theta} [\Phi_0](z) = \frac{1}{ \max_k(\beta_k)} \int_0^{e^{i\theta}\infty} \frac{\dd \zeta}{\zeta} \left(\frac{\zeta}{z} \right)^{\frac{b}{\beta} } \exp[ -\left(\frac{\zeta}{z} \right)^{\frac{1}{\beta} } ]  \mathcal{B}[\Phi_0] (\zeta) .
\end{equation}
This reproduces the asymptotic expansion $\Phi_0(t)$ for $z=te^{i\theta}$, $t\to0^+$, but now defines a function computable for all values of $t$ (and $\theta$).

Ambiguities arise if the Borel transform presents singularities along the integration path $e^{i\theta}[0,\infty]$ \footnote{Additionally, the behavior at the endpoints $\{ 0 , e^{i\theta}\infty \}$ might be of importance as well.}.
One then defines lateral summations $\mathcal{S}_\theta^{\pm}$ by deforming the contour around the singularities, the new paths we denote by $C_\theta^{\pm}$.
This indicates a branch cut of the (directional) Borel resummation at $z=t  e^{i\theta}$ with discontinuity $\left( \mathcal{S}_\theta^{-} - \mathcal{S}_\theta^{+}\right) [\Phi_0](t)$.
The discontinuity is purely non-perturbative, and it includes the exponential corrections, plus the expansions $\Phi_k(t)$ around them and the Stokes constants $S_k$ as coefficients.
Given a Borel resummable asymptotic series, the discontinuity  provides the structure of the non-perturbative terms in the trans-series and hence the Borel resummed expansions $\mathcal{S}_\theta^{\pm}[\Phi_k] (t)$ around the exponential corrections.
However, this will not remove the ambiguities in the choice of the parameters $\sigma_k^{\pm}$.
To promote the asymptotic expansion into a full trans-series we need in general to impose extra conditions.

\paragraph{Zeta function on the four sphere.}
The large charge expansion in $D=4+1$ on the four-sphere is an asymptotic series and can be computed starting from the result on the two-sphere as we have seen in Section~\ref{sec:five}.
We can therefore make use of the same resurgent techniques used in~\cite{Dondi:2021buw} to study the non-perturbative corrections to the asymptotic series.

The asymptotic expansion of the heat kernel trace
\begin{equation}
  \Tr[ e^{t \pqty{\Laplacian_{S^2} - 1/(4 r_0^2)}} ] \sim \frac{r_0^2}{t} \Phi^{S^2}(t)
\end{equation}
on the two-sphere is
\begin{align}
\Phi^{S^2}(t) &=  \sum_{n\geq 0} a^{S^2}_n \frac{t^n}{r_0^{2n}} ,  & a^{S^2}_n &=  \frac{ (1-2^{1-2n})  B_{2n} }{(-1)^{n+1}  n!} = \frac{\Gamma\big(n+\frac{1}{2}\big)}{\sqrt{\pi}} \sum_{k\neq0} \frac{(-1)^{k+1}}{(\pi k)^{2n} }
\end{align}
On the four-sphere, using Eq.\eqref{eq:RelationshipFourSphereToTwoSphere}, we find that the asymptotic expansion heat kernel trace
\begin{equation}
\Tr[ e^{t \pqty{\Laplacian_{S^4} - 9/(4 r_0^2)}} ]  \sim \frac{r_0^4}{t^2} \Phi^{S^4}(t)
\end{equation}
is given by
\begin{equation}
\Phi^{S^4}(t) = \frac{1}{6} + \sum_{\substack{n\geq 1 \\ k\neq 0 } }  \frac{(-1)^k}{\pqty{\pi k}^{2n}} \frac{1}{6\sqrt{\pi} } \left[\Gamma\big(n+\frac32\big) - \frac{3}{2}\Gamma\big(n+\frac12\big) + \frac{(\pi k)^{2}}{4} \Gamma\big(n-\frac12\big) \right] \frac{t^n }{r_0^{2n} } ,
\end{equation}
which we can use to extract the heat kernel coefficients (see also~\cite{Kluth:2019vkg}).

Comparing this to Eq.\eqref{eq:LargeOrderBehaviorOfTheCoefficientsGeneral} we see that
\begin{align}
  A_k &= (\pi k)^2  , & b_k &= \frac{3}{2} , & \beta_k &= 1  , &  \begin{cases} 
    &\frac{S_k}{2\pi i} a_{0;k} = (-1)^k \abs{k}^3 \frac{\pi^{5/2}}{6} \\
    &\frac{S_k}{2\pi i} a_{1;k} = (-1)^{k+1} \abs{k} \frac{\sqrt{\pi}}{4} \\
    &\frac{S_k}{2\pi i} a_{2;k} = (-1)^k \abs{k} \frac{\sqrt{\pi}}{24} \end{cases}  , && a_{l>2;k} &= 0 .
\end{align}
The trans-series of the four-sphere heat kernel trace therefore has to include exponentials of the form
\begin{equation}\label{eq:FormOfTheExponentialCorrectionsFourSphere}
\text{exp. corr.} \sim 2i  e^{-\frac{(\pi r_0 k)^2}{t} } \left(\frac{r_0^2 \pi}{t}\right)^{7/2} (-1)^k \abs{k} \left( \frac{k^2}{6} - \frac{ t}{4 r_0^2 \pi^2} + \frac{ t^2 }{24 r_0^4 \pi^2} \right) .
\end{equation}

To compute the exponential terms we perform the Borel resummation of the heat kernel.
On the two-sphere the Borel resummation of the asymptotic expansion is given by~\cite{Dondi:2021buw}%
\footnote{In order to avoid a Borel transform with branch cuts one performs the mapping $\zeta \to \zeta^2$ in Eq.\eqref{eq:BorelTransformGeneralDefinition}.}
\begin{align}
\mathcal{S}_0 [\Phi_0^{S^2}](t) &=  \frac{ 2 r_0 }{ \sqrt{\pi  t} }  \underset{0}{\overset{\infty}{\int}} \dd{y} \frac{y  e^{-\frac{y^2 r_0^2}{t}} }{\sin(y) } .
\end{align}
This we can use, along with Eq.\eqref{eq:RelationshipFourSphereToTwoSphere}, to directly infer that the Borel resummation of the asymptotic expansion on the four-sphere is
\begin{equation}
\begin{aligned}
\mathcal{S}_0 [\Phi_0^{S^4}](t) &= \frac{ 2 r_0^3 }{ \sqrt{\pi  t^3} }  \underset{0}{\overset{\infty}{\int}} \dd{y} \frac{y  e^{-\frac{y^2 r_0^2}{t}} }{\sin(y) } \left[ \frac{t}{4 r_0^2}- \frac{y^2}{6} - \frac{t^2}{24 r_0^4} \right] .
\end{aligned}
\end{equation}
The integral is ill-defined and has simple poles at $\zeta=k\pi$, $k\in \mathbb{Z}$.
The discontinuity is given by the Residue theorem as
\begin{equation}
\left( \mathcal{S}_0^{-} - \mathcal{S}_0^{+}\right) [\Phi](t) = 2i  \frac{t^2}{r_0^4} \left(\frac{r_0^2 \pi}{t}\right)^{\frac72} \sum_{k\neq0} (-1)^{k} \abs{k} e^{-\frac{k^2 r_0^2 \pi^2}{t}} \left[ \frac{\abs{k}^2}{6} - \frac{t}{4\pi^2 r_0^2} + \frac{t^2}{24\pi^2 r_0^4} \right] .
\end{equation}
This agrees with the form of the exponential corrections in Eq.~\eqref{eq:FormOfTheExponentialCorrectionsFourSphere}.
The Borel resummed trace of the heat kernel (with ambiguities) reads
\begin{multline}
\frac{ 2 r_0^7 }{ \sqrt{\pi  t^7} }  \underset{C_0^{\pm}}{\overset{}{\int}} \dd{y}  \frac{y  e^{-\frac{r_0^2 y^2}{t}} }{\sin(y) } \left[ \frac{t}{4 r_0^2}- \frac{y^2}{6} - \frac{t^2}{24 r_0^4} \right] \\
+  i  \left(r_0^2 \frac{ \pi}{t}\right)^{\frac72} \sum_{k\neq0}   \frac{\sigma_k^{\pm} (-1)^{k} \abs{k} }{ \pi^2} e^{-\frac{k^2 r_0^2 \pi^2}{t}} \left[ \frac{k^2 \pi^2}{3} - \frac{t}{2 r_0^2} + \frac{t^2}{12 r_0^4} \right] .
\end{multline}

It is possible to fix the ambiguities in the heat kernel trace and therefore also in all the other quantities that we will be interested in.
We can do it two ways.
Either one finds a path-integral definition where the trans-series structure arises automatically (see~\cite{Dondi:2021buw}),
or one imposes the reality of the heat kernel trace.%
\footnote{In general the reality condition  is not guaranteed to fix the non-perturbative corrections completely because there is always the possibility of $2i \sigma_k^{\pm}$ having a real part as well. But in our case it turns out to be sufficient.}
The result is
\begin{equation}
 \Tr[ e^{t \pqty{\Laplacian_{S^4} - 9/(4 r_0^2)}} ] = \frac{ 2 r_0^7 }{ \sqrt{\pi  t^7} }  \text{P.V} \underset{C_0^{\pm}}{\overset{}{\int}} \dd y  \frac{y  e^{-\frac{r_0^2 y^2}{t}} }{\sin(y) } \left[ \frac{t}{4 r_0^2}- \frac{y^2}{6} - \frac{t^2}{24 r_0^4} \right] . 
\end{equation}
To compute the lateral Borel resummation of the zeta function we directly apply the Mellin transform to the Borel resummed trace of the heat kernel.
But to change the order of integration for $s=\pm 1/2$ the integral needs to be analytically continued by extracting the first three terms of the asymptotic expansion.
We get
\begin{equation}
\begin{aligned}
\zeta^\pm\left(s   \big|  S^4  ,  m \right) = & \frac{2 r_0^{2s}}{\sqrt{\pi}  \Gamma(s)} \underset{0}{\overset{\infty}{\int}} \dd t \,  \underset{C_0^{\pm}}{\overset{}{\int}} \dd y \,  \frac{e^{- m^2 r_0^2 t - \frac{y^2}{t}} }{t^{\frac92 - s}}  \bigg[  \frac{y }{\sin(y) } \bigg( - \frac{y^2}{6} + \frac{t}{4} - \frac{t^2}{24} \bigg) - \frac{y^2}{3} + \frac{y^4}{18} \\
& \quad \quad \quad \quad \quad \quad  + \frac{17 y^6}{5400} \bigg] + \frac{1}{6} \frac{r_0^4 m^{4-2s}}{(s-1)(s-2)} - \frac{1}{24}\frac{r_0^2 m^{2-2s}}{(s-1)} - \frac{17}{2880} m^{-2s} . 
\end{aligned}
\end{equation}
We exchange the order of integration for $s=-1/2$. The result can be identified with the (non-standard) Borel resummation of the asymptotic series $\Omega_0(m)/(2N)$ of the grand potential $\Omega(m)/(2N)$. We have
\begin{multline}
\mathcal{S}_0^\pm \Big[\frac{\Omega_0}{2N} \Big] (m) =  - \frac{r_0^3 m^4}{24\pi} \underset{C_0^{\pm}}{\overset{}{\int}} \frac{\dd y}{y^2}\, \bigg[\frac{y }{\sin(y) } \bigg( 2\Big[ K_4 (2r_0 my)  + K_{2}(2r_0 my) \Big] + \frac{ K_2 (2r_0 my)}{(r_0 m)^2}  \bigg) \\+ \bigg( 8 
- \frac{4y^2}{3 } - \frac{17 y^4 }{225}\bigg) K_4 (2r_0 my) \bigg]  + \frac{2}{45} r_0^4 m^{5} + \frac{1}{36} r_0^2 m^{3} - \frac{17}{2880}m  ,
\end{multline}
where $K_n(x)$ is the modified Bessel function of the second kind of order $n$. The discontinuity is given by
\begin{multline}
\left( \mathcal{S}_0^{-} - \mathcal{S}_0^{+}\right) \Big[\frac{\Omega_0}{2N} \Big] (m) = i \frac{r_0^3 m^4}{12} \sum_{k\neq 0}  \frac{(-1)^{k}}{ |k|} \bigg( 2 \Big[ K_{4} (2\pi r_0  mk)  + K_{2}(2\pi r_0 mk) \Big] \\
  + \frac{1}{ (r_0 m)^2} K_{2}(2\pi r_0 mk) \bigg) .
\end{multline}
For our purposes it is sufficient to only consider the leading non-perturbative terms appearing in the free energy. To do so we will use Hankel's asymptotic expansion for the modified Bessel functions of the second kind
\begin{align}
K_\alpha (z) &\sim \sqrt{ \frac{\pi}{2z} } e^{-z} \left( 1 + \frac{(4\alpha^2-1)}{8z} \left( 1 + \frac{1}{2}\frac{(4\alpha^2- 3^2)}{8z} \left( 1+ \frac{1}{3} \frac{(4\alpha^2-5^2)}{8z} \Big(1 + \dots \quad \Big) \right) \right) \right)  ,
\end{align}
for $-3\pi/2 < \arg(z) < 3\pi/2$. The first non-perturbative exponential terms are
\begin{equation}
\omega(m) \supset i \frac{r_0^3 m^4}{12}  \frac{(-1)^{k}}{ |k|} \bigg(  4  + \frac{1}{(r_0 m)^{2}}  \bigg) \frac{1}{\sqrt{4 r_0 m|k|} } e^{-2\pi r_0 m|k|} \sim i \frac{(r_0 m)^{\frac72}}{6 r_0}  \frac{(-1)^{k}}{ |k|^{\frac32}} e^{-2\pi r_0 m|k|}    .
\end{equation}
If we use the first term of the order-by-order Legendre transform of the perturbative part, i.e. $q = - (r_0m)^4/9$, $r_0 f(q)=r_0 mq + (r_0m)^5/45$, with $q=Q/2N$ , then the leading correction to the free energy per \ac{dof} is
\begin{equation}
r_0 f(q) \supset i \frac{\tilde q^{7/8}}{6 \sqrt{f_1} }  \frac{(-1)^{k}}{ |k|^{\frac32}} e^{-2\pi f_1 \tilde q^{1/4} |k|} ,
\end{equation}
where $\tilde q = 9q$ and $f_1$ is the complex phase coming from the negative sign in the first equation and the fact that one has to pick a branch for the root. Finally, following the reality prescription for the heat kernel trace, we can put the grand potential into the form
\begin{equation}
\omega(m) = - \frac{r_0^3 m^4}{24\pi}  \text{P.V.}\underset{0}{\overset{\infty}{\int}} \frac{ \dd y }{y \sin(y) } \bigg( 2\Big[ K_4 (2my)  + K_{2}(2my) \Big] + \frac{1}{(r_0 m)^2} K_2 (2my) \bigg)  . 
\end{equation}

\paragraph{Optimal truncation.}%
An alternative approach to get a meaningful result from an asymptotic series is to truncate it.
A commonly applied rule of thumb to find a finite sum that is as close as possible to the ''actual'' value is to truncate at the term that gives the smallest contribution.
For an asymptotic series $\sum a_n x^n$ , if we suppose that the coefficients $a_n$ diverge as $(\beta n)!A^{-n}$, then the optimal truncation is at 
\begin{equation}
  N(x) \approx \frac{1}{\beta}  \abs{Ax}^{1/\beta} ,
\end{equation}
with error $\epsilon(x) \sim \exponential(-(Ax)^{1/\beta})$. For the special case $x \sim 1$ we can simply look at the ratio of consecutive coefficients of the terms in the asymptotic expansion.
Once this ratio exceeds one we truncate the series.
For the zeta function $\zeta (s \mid S^4 , m )$ in terms of the variable $\tilde m^2 = r_0^2m^2 - \sfrac94 \sim 1$ the optimal truncation is after the third term for $s=\sfrac32$, after the fourth for $s=\sfrac12$ and after the fifth for $s=-\sfrac12$.

To go beyond that we need to rely on our resurgent analysis.
For the trace of the heat kernel, using the large order behavior of the Bernoulli numbers, given by \(\frac{ (1-2^{1-2n})  B_{2n} }{(-1)^{n+1}  n!} \sim \frac{2}{\sqrt{n\pi}} \frac{n!}{\pi^{2n}}\), one finds that $\beta=1$ and $A=\pi^2$, and hence
\begin{align}
&N(t) \approx \pi^2 r_0^{-2} t , \quad  &\epsilon(t) \sim \exp(-\pi^2 r_0^{-2} t ) .
\end{align}
The asymptotic expansion of the grand potential is found by Mellin transforming the asymptotic expansion of the heat kernel trace
\begin{align}
\zeta\left(s   \big|  S^4  ,  m \right) &\sim r_0^{2s} (\tilde m^2)^{2-s} \sum_{n\geq 0}  a_n^{S^4} \frac{\Gamma(n+s-2)}{\Gamma(s)} (\tilde m^2)^{-n}   , \\
\frac{\Omega (m)}{2N} &\sim r_0^{-1} (\tilde m^2)^{\frac52} \sum_{n\geq 0}  a_n^{S^4} \frac{\Gamma\big(n-\frac52\big)}{4\sqrt{\pi} } (\tilde m^2)^{-n} = - \frac{ (\tilde m^2)^{\frac52}}{45 r_0} + \frac{(\tilde m^2)^{\frac52}}{r_0} \sum_{n\geq 1}  \frac{\Omega_n}{2N} (\tilde m^2)^{-n} ,
\end{align}
where $\tilde m^2 = r_0^2m^2 - \sfrac94$. The coefficients $\Omega_n/(2N)$ in closed form are written as ($n\geq1$)
\begin{equation}\label{eq:CoefficientsGrandPotentialWithDoubleFactorial}
\frac{\Omega_n}{2N} = \sum_{k\neq 0 } \frac{(-1)^k}{\pqty{\pi k}^{2n + 3}} \frac{(\pi \abs{k})^3}{24 \pi } \Gamma\big(n-\frac{5}{2}\big) \left[\Gamma\big(n+\frac{3}{2}\big) - \frac{3}{2}\Gamma\big(n+\frac{1}{2}\big) + \frac{(\pi k)^{2}}{4} \Gamma\big(n-\frac{1}{2}\big) \right] .
\end{equation}
The double factorial first needs to be resolved in order to match with the general higher-order trans-series coefficients, but already makes it clear that the coefficients grow like $(2n)!$.
Once the double factorial structure is resolved, then the leading term is
\begin{equation}
\frac{\Omega_n}{2N} = \sum_{k\neq 0 } \frac{(-1)^k}{\left(2\pi \abs{k}\right)^{2n-\frac{1}{2}} }\frac{1 }{12 \pi \sqrt{ \abs{k} }} \pqty{ \Gamma( 2n  - \sfrac{1}{2} )  + \dots }.
\end{equation}
What we can read off already is that
\begin{align}
A_k &= (2\pi k) , & b_k &= -\frac{1}{2} , & \beta_k &= 2 , & \frac{S_k}{2\pi i} a_{0;k} &= \frac{(-1)^k}{24 \pi \sqrt{ \abs{k} } } .
\end{align}
This tells us that the optimal truncation is
\begin{align}
N(\mu) &\approx \frac{1}{2} \abs{2\pi \tilde m^2}^{1/2} ,&  \epsilon(\tilde m) &\sim \exp(-\sqrt{2\pi \tilde m^2}) .
\end{align}
For $r_0^2 m^2 \gtrsim 3/2$, i.e. $\tilde m^2 \gtrsim 3/2$, the optimal truncation is at the first term, so that $\Omega(m)/(2N) \sim - \frac{ (\tilde m^2)^{\frac52}}{45 r_0}$. This way we recover the maximum of the grand potential $\Omega(\sfrac32 )/(2N)=0$ at $m=3/2$ from its asymptotic expansion using the optimal truncation.%
\footnote{To be precise we cannot see that it is a maximum since $\Omega(\sfrac32 )'/(2N) \sim - (r_0 m) \frac{ (\tilde m^2)^{\frac32}}{9}\big|_{m=3/2} = 0$.}

\paragraph{Flex from large charge.}
The second derivative of the grand potential reads (we use the identity $\zeta' (s \mid \mathcal{M} , m) = - 2ms \zeta (s+1 \mid \mathcal{M} , m)$ here)
\begin{equation}
\frac{\Omega''(m)}{2N} = - \frac{1}{2} \pqty{ \zeta(\sfrac{1}{2}   \mid  S^4 , m ) - m^2  \zeta(\sfrac{3}{2}   \mid  S^4 , m )} .
\end{equation}
It is clear that the asymptotic expansion in terms of $\tilde m^2 = r_0^2 m^2 - 9/4$ cannot reproduce the flex at $m\approx 1.266<3/2$.
So we further expand in an asymptotic expansion in terms of $m$. Around $m\sim1$ the optimal truncation can be found by looking again at the ration of consecutive coefficients.
This ratio starts exceeding the value one after the sixth term.
Up to this term the grand potential is
\begin{equation}
\frac{\Omega(m)}{2N} = - \frac{r_0^4m^5 }{45}  + \frac{r_0^2 m^3}{9} - \frac{29 m}{45} + \frac{37 m^{-1}}{756r_0^2} + \frac{149 m^{-3}}{15120r_0^4} + \frac{179 m^{-5}}{55440 r_0^6 } . 
\end{equation}
Expanded up to the sixth term the second derivative of the grand potential reads
\begin{equation}
\frac{\Omega''(m)}{2N} = -\frac{4 r_0^4 m^3}{9} + \frac{2 r_0^2 m}{3} + \frac{37 m^{-3}}{378 r_0^2} + \frac{149 m^{-5}}{1260 r_0^4} + \frac{179 m^{-7}}{1848 r_0^6 } . 
\end{equation}
The $m^{-1}$ term is of course missing and it also cancels exactly between the two zeta functions.
\begin{figure}
  \centering
  \includegraphics[width=.55\textwidth]{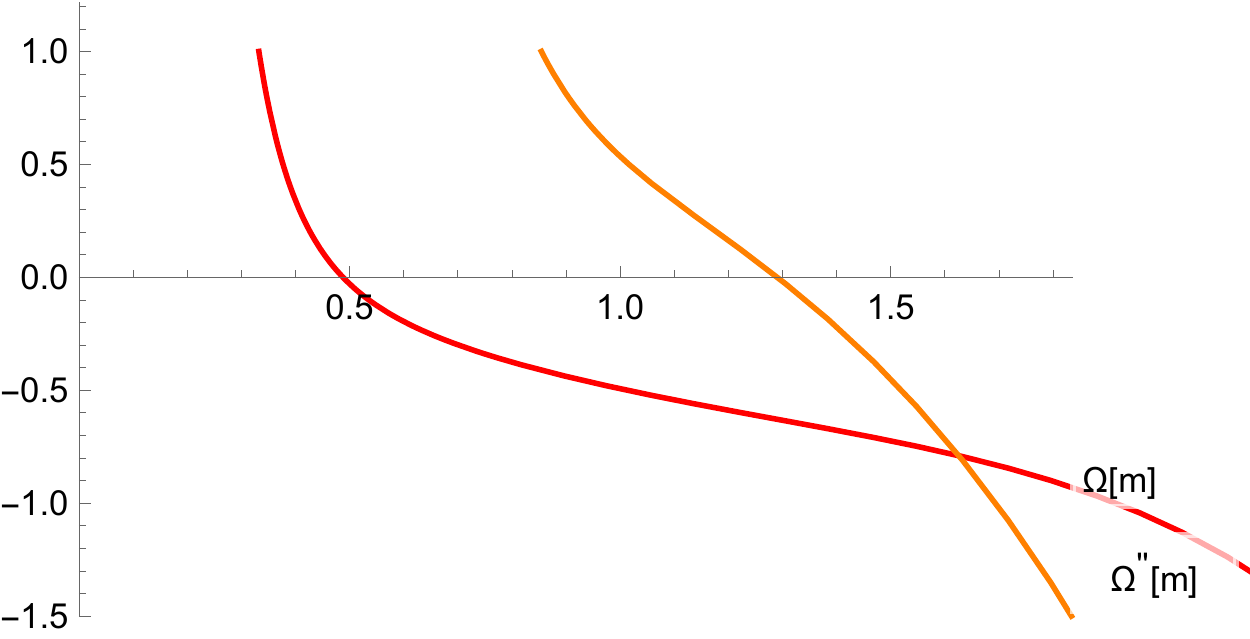}
  \caption{The truncated asymptotic expansion of the grand potential \({\Omega(m)}\) (red) and its second derivative \({\Omega''(m)}\) (orange) on the four-sphere.There is a flex for $m\approx1.290$.
  }
  \label{fig:S4-omegaasymptotic}
\end{figure}

The asymptotic expansions of ${\Omega(m)}$ and ${\Omega''(m)}$ are shown in Figure~\ref{fig:S4-omegaasymptotic}.
Numerically one finds that the asymptotic expansion of ${\Omega''(m)}$ has a zero at $r_0 m\approx1.290$, i.e. the asymptotic expansion of $\omega(m)$ has a flex at $r_0 m\approx1.290$. 
This is very close to the actual value of $r_0 m_{\text{fl}} \approx 1.266$. 
Looking at the graph it is clear though that the asymptotic expansion does not reproduce the maximum of the grand potential at $r_0 m = 3/2$.
The error of the optimal truncation at $\tilde m^2 \approx 1$ is given by $\epsilon(\tilde m^2 \sim 1) \sim = e^{ - \sqrt{2\pi} } = 0.08$.

\setstretch{1}

\printbibliography{}

\end{document}